\documentclass{aa}  

\usepackage{graphicx}
\usepackage{txfonts}
\begin{document}

   \title{The role of mass loss in constraining quenching time in dwarf galaxies from AGB and RGB star counts}
\author{P. Ventura\inst{1}, R. D'Souza\inst{2}, F. Dell'Agli\inst{1}, 
E. F. Bell\inst{3}, C. Gavetti\inst{1,4}, C. Fiumi\inst{4}, M. Tailo\inst{5}}

   \institute{INAF, Observatory of Rome, Via Frascati 33, 00077 Monte Porzio Catone (RM), Italy \\ email: paolo.ventura@inaf.it \and 
   Vatican Observatory, 00120 Vatican City State \and
   Department of Astronomy, University of Michigan, 1085 S. University Ave, Ann Arbor, MI 48109-1107, USA \and
   Dipartimento di Matematica e Fisica, Università degli Studi Roma Tre, 
   via della Vasca Navale 84, 00100, Roma, Italy \and
   Dipartimento di Fisica e Astronomia Augusto Righi, Università degli Studi di Bologna, Via Gobetti 93/2, 40129 Bologna, Italy 
   }

   \date{2023}

 
  \abstract
   {The capability of reconstructing the past star formation history of dwarf elliptical galaxies 
   out of the Local Volume relies on modelling bright stellar populations currently
   evolving through the red giant branch (RGB) and the asymptotic giant branch (AGB) phases.
   Recent studies proposed the use of the relative fractions of RGB and AGB stars populating 
   specific boxes in the observational colour-magnitude plane to infer the epoch within
   which $90\%$ of the stellar population of the galaxy formed (T90).}
   {We aim at understanding the physical process of stellar evolution that are constrained by 
   the relationship between the relative fraction of AGB and RGB stars of dwarf galaxies and 
   the T90 epoch.}
   {We use updated stellar models that include the description of dust formation in the
   wind, to undertake a population synthesis approach, to allow monitoring the variation of the
   $\rm N_{AGB}/N_{RGB}$ ratio with time. The effects of some specific ingredients, such as
   the mass loss experienced by low-mass stars during the RGB phase, and the details of the
   time variation of the star formation rate, are extensively explored and tested against
   data.}
   {The mass lost by low-mass stars during the RGB evolution proves the most
   relevant ingredients affecting the time variation of $\rm N_{AGB}/N_{RGB}$: at
   metallicities $\sim 1/10$ solar, a mass
   loss $\rm \sim 0.25~M_{\odot}$ is required to reproduce the observations. This
   analysis allows to derive a relationship between $\rm N_{AGB}/N_{RGB}$ and T90,
   with a $\sim 1$ Gyr uncertainty on T90.
   }
   {}


   \titlerunning{t$_{90}$ and its future applications}
   \authorrunning{P. Ventura et al.}
   \maketitle
%

\section{Introduction}

Our ability to reconstruct the star formation history (SFH) of nearby individual galaxies 
by resolving their stellar populations has helped build physical intuition into the processes 
that form and shape galaxies -- from reconstructing how the stellar populations in the solar neighborhood grew over time \citep{Gallart2024} and have been influenced by recurrent impacts of the Sagittarius dwarf galaxy \citep{Ruiz-Lara2020}, to reconstructing how the nearby Andromeda galaxy suffered a global burst of star formation 2 Gyr ago \citep{Williams2015} due to a merger of a large progenitor \citep{richard18b,Hammer2018} leading to the thickening of its stellar 
disk \citep{Dorman2015, Dalcanton2023}. Further out, constraints on the quenching of star 
formation of the stellar halos of nearby MW-mass galaxies \citep[e.g., M31, Cen A and M81; ][]{Brown2006, Rejkuba2005, Durrell2010} have given us useful constraints about the time 
of merger of their most massive progenitors \citep{deason15,richard18,monachesi19,wang20}.  
At smaller scales, constraints of the quenching of star formation in low-mass dwarf 
galaxies have been useful in 
constraining their infall time into the MW \citep[e.g.,][]{Fillingham2015, Slater2014}, 
while similar information in ultra-faint MW and M31 dwarfs have given us useful constraints 
about the epoch and the progress of reionization \citep{Weisz14}. While some of these insights critically 
depend on the temporal resolution of the SFH obtained from faint but more numerous main 
sequence turnoff stars, others have been gleaned from brighter red clump/horizontal 
branch stars which offers us the ability to study a larger number of galactic systems out 
to 3.5 Mpc albeit at lower temporal resolution \citep{Weisz2011}. If the community wishes 
to extend this intuition to the abundance of galactic systems found further out within 
the Local Volume (<10 Mpc), then it needs to make a concerted effort to understand how 
to constrain the SFH from brighter red giant branch (RGB) and asymptotic giant branch (AGB) 
stars readily found in HST and JWST observations \citep[e.g. ][]{Rejkuba2022}.

Our ability to capitalize on the temporal information present in the longer-lived RGB 
stars ($10^8-10^9 \mathrm{yrs}$, depending on the stellar mass) and the short-lived ($\sim10^5 \mathrm{yrs}$) intermediate-age AGB stars depends 
critically on our ability to model the latter. Due to their extremely short life time, 
AGB stars are difficult to find in open and globular clusters, and hence earlier single 
stellar population models have had a hard time constraining their behaviour. 
Recent efforts have focused
on calibrating models against optical and NIR photometry and spectroscopy of AGB populations
in nearby galaxies that show recent star formation including M31, Magellanic Clouds, 
dwarf irregulars and nearby spirals with great success \citep[e.g.,][]{Girardi2010, 
Rosenfeld2014, Choi2016, Pastorelli2020, flavia2016,  flavia2018, flavia2019, cla}.

\begin{figure}
\begin{minipage}{0.48\textwidth}
\resizebox{1.\hsize}{!}{\includegraphics{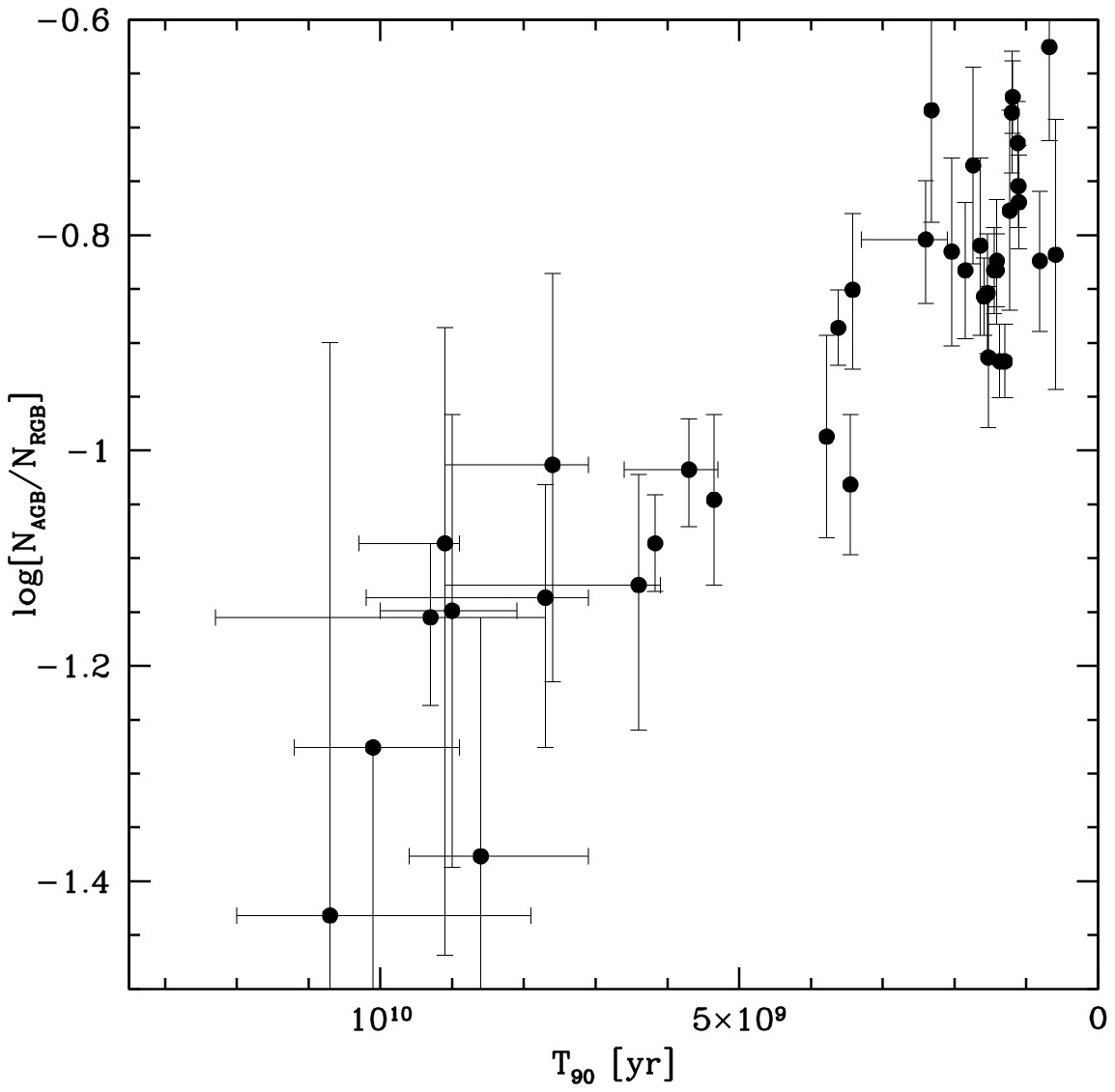}}
\end{minipage}
\vskip-60pt
\caption{The updated correlation between the relative fraction of the number 
of AGB and RGB stars ($\rm N_{AGB}/N_{RGB}$) in nearby dwarf elliptical galaxies and
the time before which $90\%$ of the stellar population of a given galaxy 
formed, $\rm T_{90}$ \citep[][update in Appendix \ref{appendix:a1}]{harmsen23}.}
\label{figdata}
\end{figure}

However, the behaviour of low-mass low-metallicity AGB stars in dwarf elliptical galaxies, 
whose star formation has been quenched a long time ago, has not been sufficiently explored 
until now, partly due to the low number of AGB stars found in these galaxies. Recently, 
\citet[][hereafter H23]{harmsen23} demonstrated that the relative fraction of the number 
of AGB and RGB stars ($\rm N_{AGB}/N_{RGB}$) in nearby dwarf elliptical and irregular galaxies 
is correlated  with the quenching time of star formation. The results by H23 are
shown in Fig.~\ref{figdata}, in terms of the T90 vs $\rm N_{AGB}/N_{RGB}$ trend.
In practice, H23 used HST resolved observations of Local Group dwarfs in the $\rm F606W$ and $\rm F814W$ filters to derive 
a correlation between the time before which $90\%$ of the stellar population of a given galaxy 
formed, $\rm T_{90}$, and the $\rm N_{AGB}/N_{RGB}$ ratio. The latter quantity is estimated by 
counting the sources populating specific regions in the $\rm (F606W-F814W, F814W)$ diagram, 
chosen to map the stellar population just below the tip of the red giant branch (TRGB), 
and those $0.15$ mag brighter than the TRGB, which can be safely assumed to evolve through 
the AGB phase. A unique feature of the H23 sample is the inclusion of a number of 
M31's dwarf elliptical galaxies whose star formation quenched more than 5-6 Gyr ago, 
suggesting that AGB O-stars from low-mass stars may be present in galaxies quenched long ago. 
The recent publication of the SFH of all the dwarf galaxies of M31 by \citet{savino25} 
has only reinforced this relation (see Appendix \ref{appendix:a1}). Although H23 used this relation
to determine the quenching time of the stellar haloes of nearby Milky Way-mass galaxies in the
GHOSTS sample, they did not comment on what particular physical processes of the evolution 
of AGB stars are constrained by this relation.

Against this background, we embarked on consolidating the methodology
proposed by H23 and understanding the physical processes involved by applying results 
from detailed stellar evolution modelling coupled to the description of dust 
formation in the wind, to understand how the relative fraction of AGB and RGB stars in a 
given galaxy depends on the SFH. Although the time scale of the RGB evolution 
of stars can be determined with sufficient accuracy, the time variation of 
the post core-helium burning phases, a key factor in determining the expected 
number of AGB sources in a given region of any colour-magnitude plane, is 
sensitive to various physical mechanisms, which deeply affect the 
evolution through the AGB. The most relevant in this regard are: 
a) the mass loss experienced by the stars during the ascending of the 
RGB, which might lead them to skipping the AGB phase completely; b) the 
efficiency of the mechanisms favouring the alteration of the surface 
chemistry, which might importantly change the strength 
of the AGB mass loss, thus affecting the evolutionary time scales; c) the 
formation of dust in the circumstellar envelope, which would trigger 
a shift of the spectral energy distribution (SED) towards the infrared 
region.

In this work, we focus on stellar populations with metallicity $\rm [Fe/H]=-1$,
which reflects the chemistry of most of the dwarf galaxies investigated by H23. 
After a preliminary analysis of the various factors that affect the 
description of the AGB phase, we discuss in detail a few selected galaxies, 
in order to span  a variety of SFHs, ranging from those characterised by very old
stellar populations, such as Sculptor and NGC 185, to those believed
to have experienced recent star formation, such as Fornax. 
The results of the analysis applied to the selected galaxies are then
used in a more general approach to discuss the possibility of
determining a relationship between the relative consistency of the
RGB and AGB stellar population and the SFH.
In the final part of the paper we propose 
moving from the optical filters employed by H23 to the near-infrared filters employed by JWST, Euclid and WFIRST.

\section{RGB and AGB evolution}
\label{evol}
The methodology introduced by H23 to reconstruct the time of shutdown in the star 
formation of galaxies is based on the relative fractions of star counts in the RGB 
and AGB boxes defined in the $\rm (F606W-F814W, F814W)$ diagram. The corners of the 
RGB selection box (relative to the apparent magnitude of the TRGB defined in VEGA 
magnitudes) are located at 
$(0.8, 0), (1.9, 0), (0.68, 0.6), (1.78, 0.6)$, while the choice for the AGB box 
was $(1.0, -0.75)$, $(2.2, -0.75)$, $(0.88, -0.15)$, $(2.08, -0.15)$. 
To discuss the potentialities of the $\rm N_{AGB}/N_{RGB}$ ratio as an indicator 
of the quenching time of star formation of galaxies, it is important to understand
how stars of different masses pass in $\rm F814W$ magnitude within the two boxes.
Hence, we consider the evolutionary tracks of stars of different masses with 
chemical composition $\rm [Fe/H]=-1$, which is the metallicity of galaxies addressed 
in this work. It is important to emphasize that the colour limits of the boxes serve
to observationally separate various phases of stellar populations from each other as 
well as from contaminants and interlopers.

\subsection{The evolution of stars in the RGB and AGB boxes}
The evolutionary sequences used in the present analysis are
based on stellar models calculated by means of the ATON code
for stellar evolution \citep{italo}, interfaced with the description of the
dust formation process in the wind, according to the schematization
discussed in \citet{ventura12}. We adopt the $\rm [Fe/H]=-1$ models published 
in \citet{ventura14}, recently extended to the post-AGB phase by \citet{devika23}. 
In the following, we will refer to these models as the ATON models.
All the evolutionary sequences adopted are available at CDS.

\begin{figure}
\begin{minipage}{0.48\textwidth}
\resizebox{1.\hsize}{!}{\includegraphics{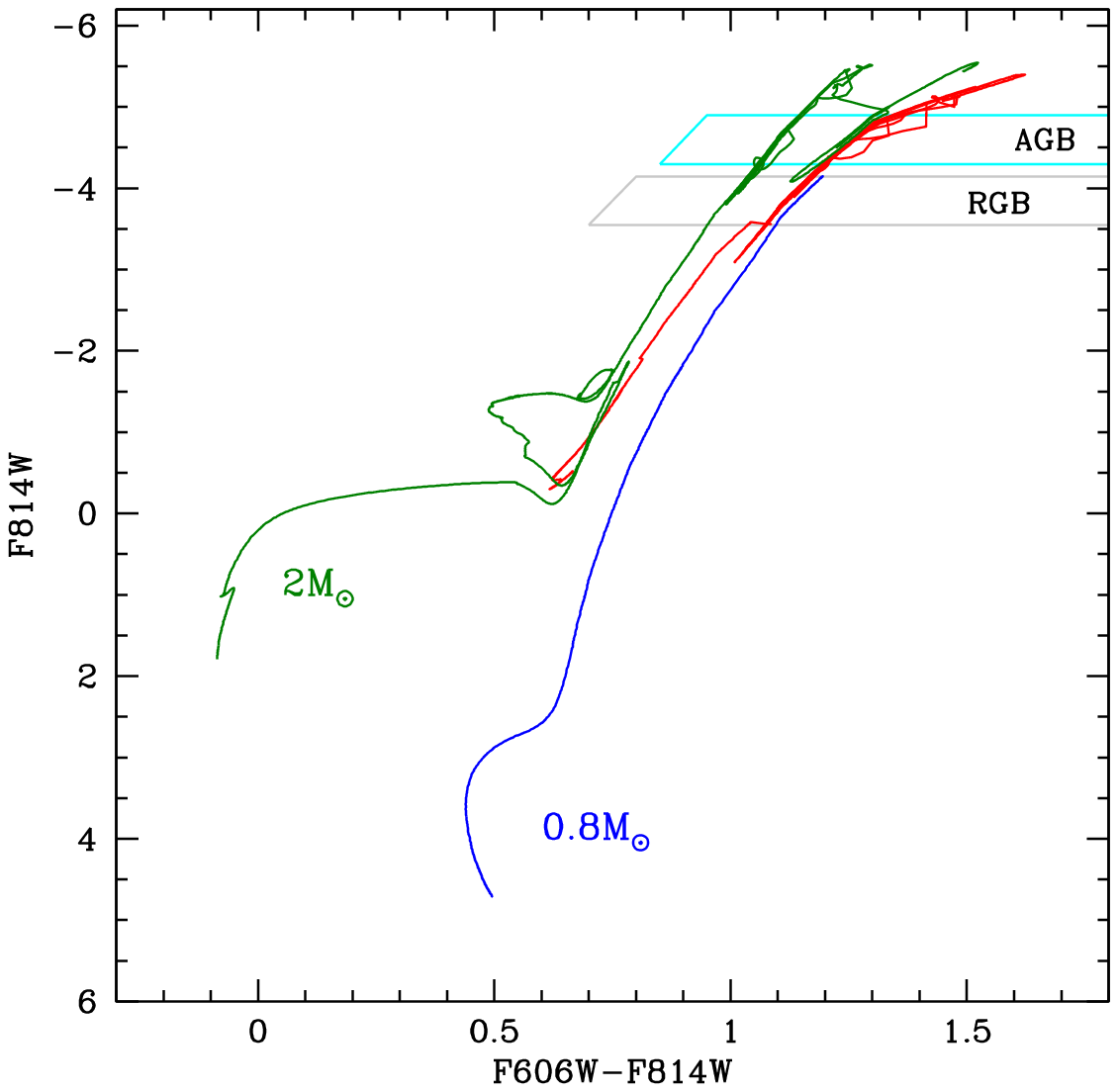}}
\end{minipage}
\vskip-60pt
\caption{Evolutionary tracks of model stars of metallicity 
$\rm [Fe/H]=-1$ and initial masses $\rm 0.8~M_{\odot}$
(blue line, until the helium-flash, then red line, from
the start of the core helium burning phase) and $\rm 2~M_{\odot}$ 
(green), on the colour-magnitude $\rm (F606W-F814W, F814W)$ diagram.
The regions inside the grey and cyan lines indicate the RGB and
AGB boxes, respectively, introduced by H23. The evolutionary
sequences are shown until the start of the post-AGB contraction,
which is omitted for readability.
}
\label{fcmd}
\end{figure}

\begin{figure*}
\begin{minipage}{0.48\textwidth}
\resizebox{1.\hsize}{!}{\includegraphics{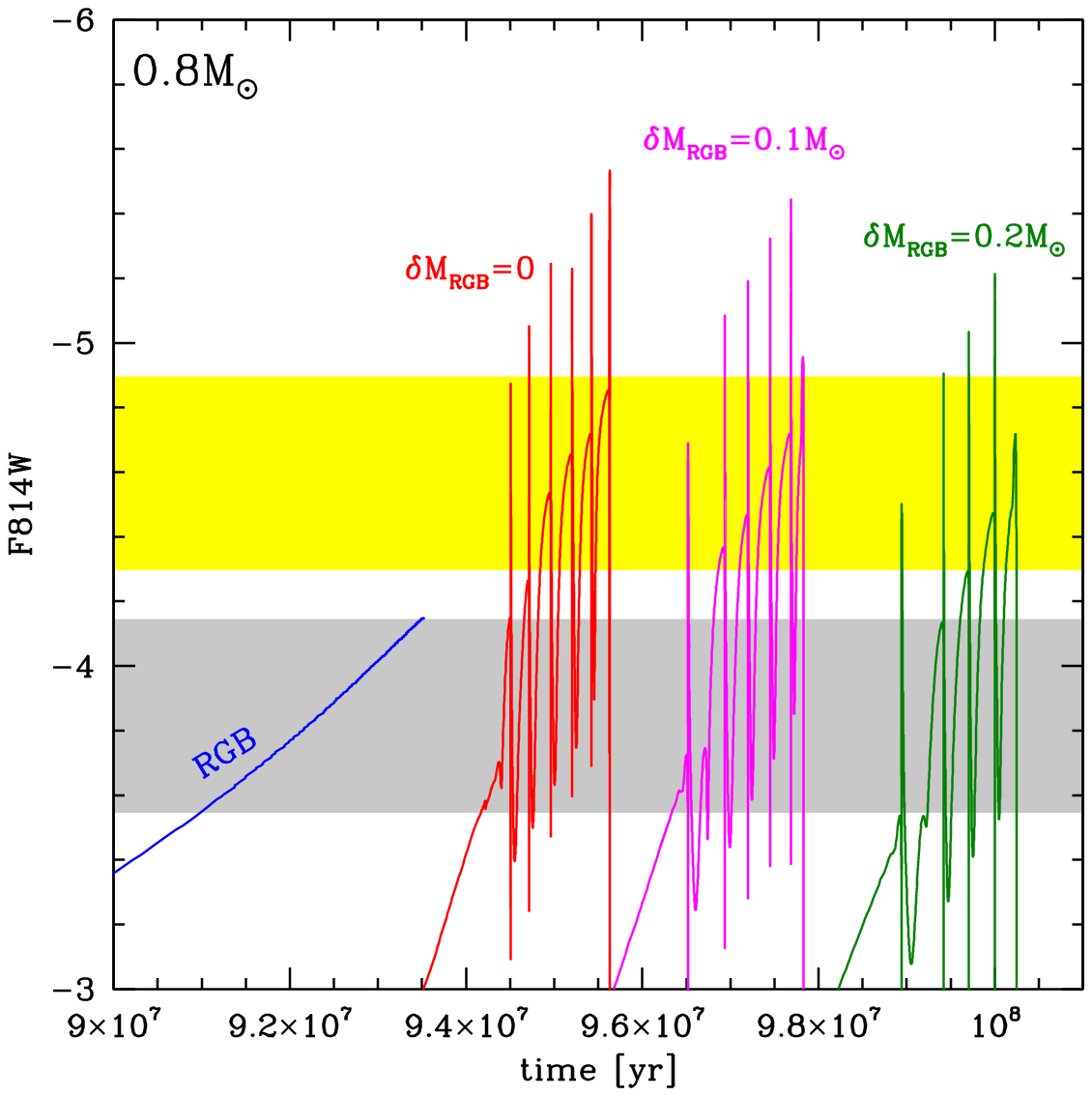}}
\end{minipage}
\begin{minipage}{0.48\textwidth}
\resizebox{1.\hsize}{!}{\includegraphics{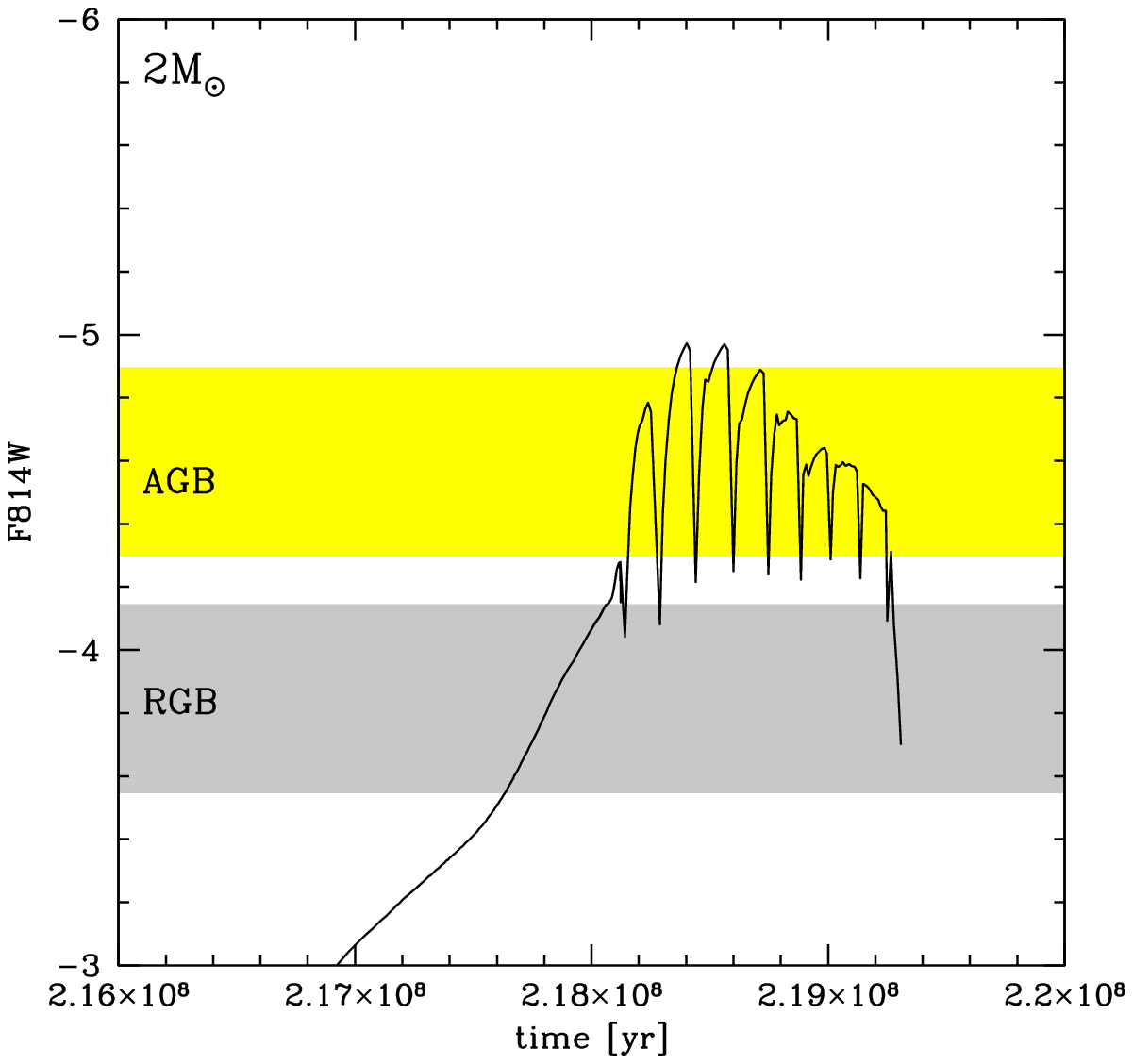}}
\end{minipage}
\vskip-90pt
\begin{minipage}{0.48\textwidth}
\resizebox{1.\hsize}{!}{\includegraphics{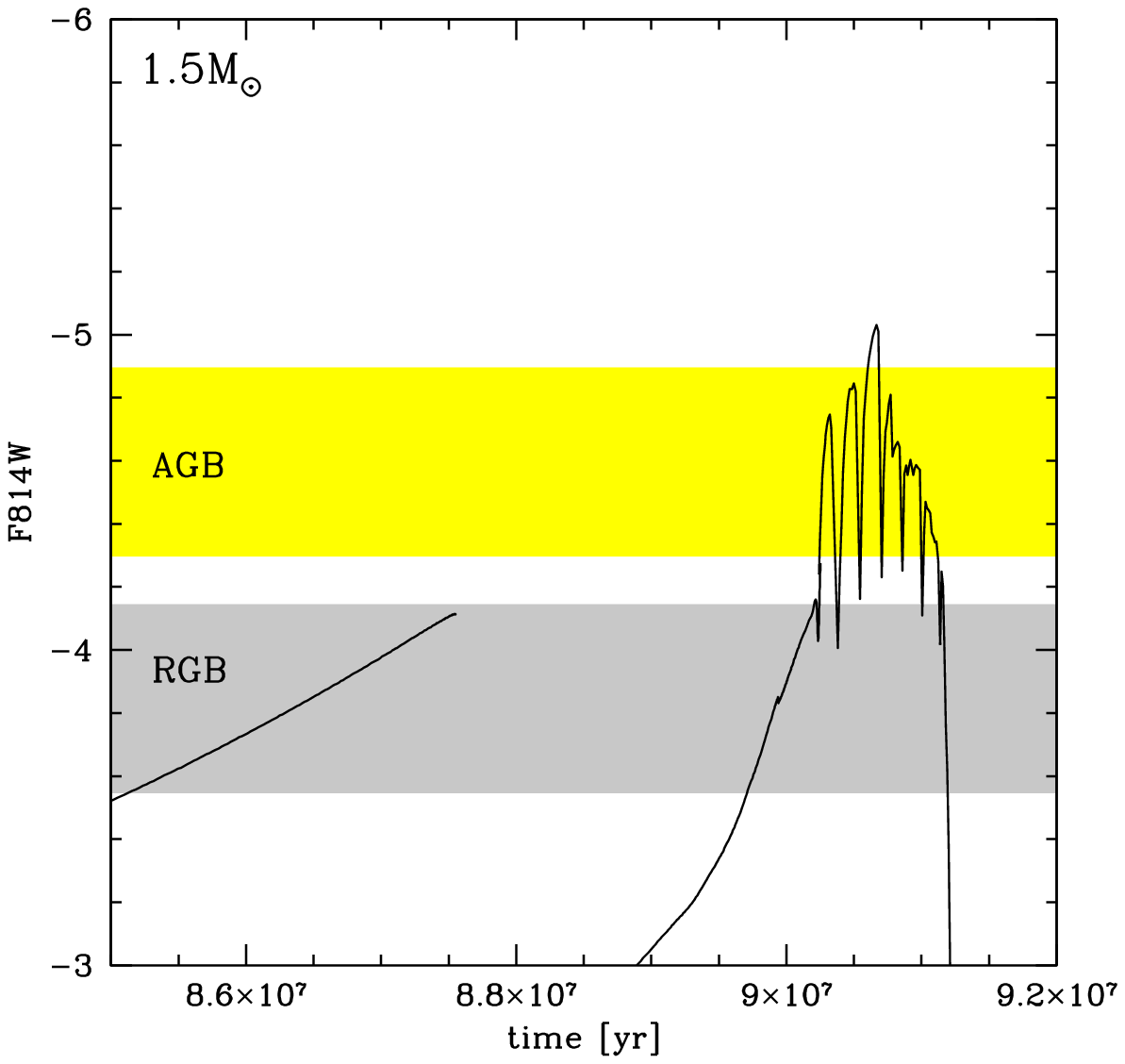}}
\end{minipage}
\begin{minipage}{0.48\textwidth}
\resizebox{1.\hsize}{!}{\includegraphics{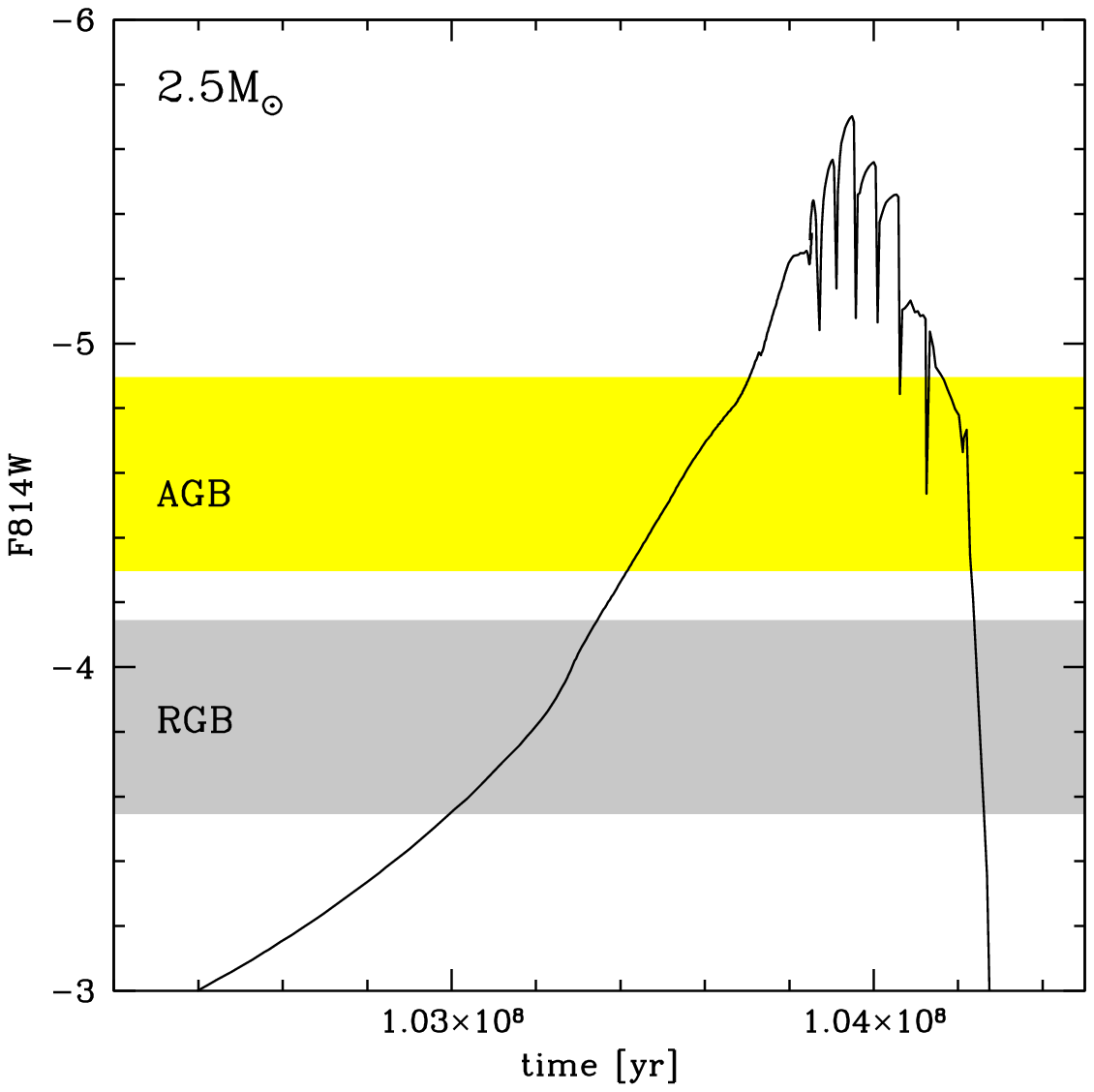}}
\end{minipage}
\vskip-50pt
\caption{Upper panels: Time variation of the F814W magnitude of the 
$\rm 0.8~M_{\odot}$ and $\rm 2~M_{\odot}$ model stars shown in
Fig.~\ref{fcmd}. The grey and yellow regions refer to the
F814W magnitudes of the RGB and AGB boxes introduced by H23.
The evolution of the $\rm 0.8~M_{\odot}$ model star is divided into
the RGB (blue line) and the post-flash (red) parts. The magenta
and green lines refer to the evolution of the $\rm 0.8~M_{\odot}$ 
model star, under the hypothesis that $\rm 0.1~M_{\odot}$ and
$\rm 0.2~M_{\odot}$ were lost during the RGB phase. Lower panels: Time 
variation of the F814W magnitude of $\rm 1.5~M_{\odot}$ and 
$\rm 2.5~M_{\odot}$ stars.}
\label{f0820}
\end{figure*}

Fig.~\ref{fcmd} shows the evolutionary tracks of stars of initial mass 
$\rm 0.8~M_{\odot}$ and $\rm 2~M_{\odot}$ in the diagram $\rm (F606W-F814W, F814W)$. 
The former experiences the 
helium flash, while the latter undergoes quiescent core helium burning. The 
evolution of the $\rm F814W$ flux of the two model stars considered 
is shown in the top panels of Fig.~\ref{f0820}. In this subsection we consider 
the case of zero mass loss during the RGB evolution of low-mass stars.  
The effects of the RGB mass loss will be addressed in Section \ref{mloss}.

The evolutionary track of the $\rm 0.8~M_{\odot}$ model star enters the RGB 
box, delimited by the grey lines in Fig.~\ref{fcmd}, which corresponds to the 
grey area in Fig.~\ref{f0820}, twice. The first 
crossing occurs during the final RGB phases, 
before the TRGB is reached; the second 
overlapping takes place during the
early AGB evolution and the first two inter-pulse phases.
The determination of $\rm N_{RGB}$ is more affected by the
first crossing, which is $\sim$ four times longer than the
second. The AGB box, delimited by cyan lines in
Fig.~\ref{fcmd}, and indicated with yellow shading
in Fig.~\ref{f0820}, is crossed only during the 
AGB evolution, for the last four thermal pulse (TP) phases. 

The $\rm 2~M_{\odot}$ model star evolves through the RGB box at the start 
of the AGB evolution, the so-called early-AGB phase, before the ignition of
the TPs. The overlap of the evolutionary track and the AGB box takes 
place during all the inter-pulse phases of the TP-AGB evolution, until the envelope is lost.
In the top panels of Fig.~\ref{f0820} we see that both the model stars
considered eventually evolve into the RGB box towards the very final evolutionary
phases; however, this late crossing of the RGB box is so short that is not 
relevant for the determination of $\rm N_{AGB}/N_{RGB}$.

\begin{figure*}
\begin{minipage}{0.48\textwidth}
\resizebox{1.\hsize}{!}{\includegraphics{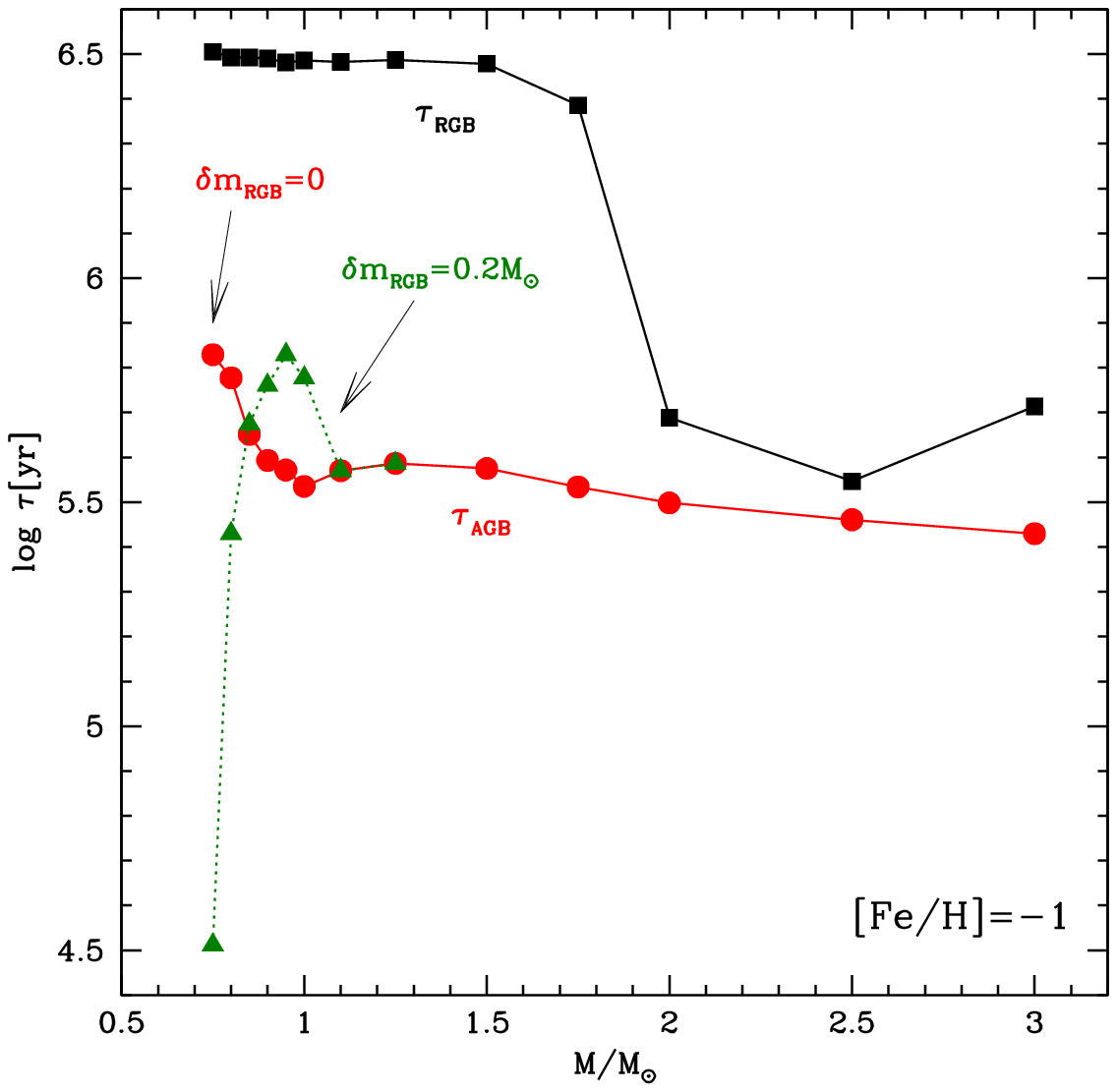}}
\end{minipage}
\begin{minipage}{0.48\textwidth}
\resizebox{1.\hsize}{!}{\includegraphics{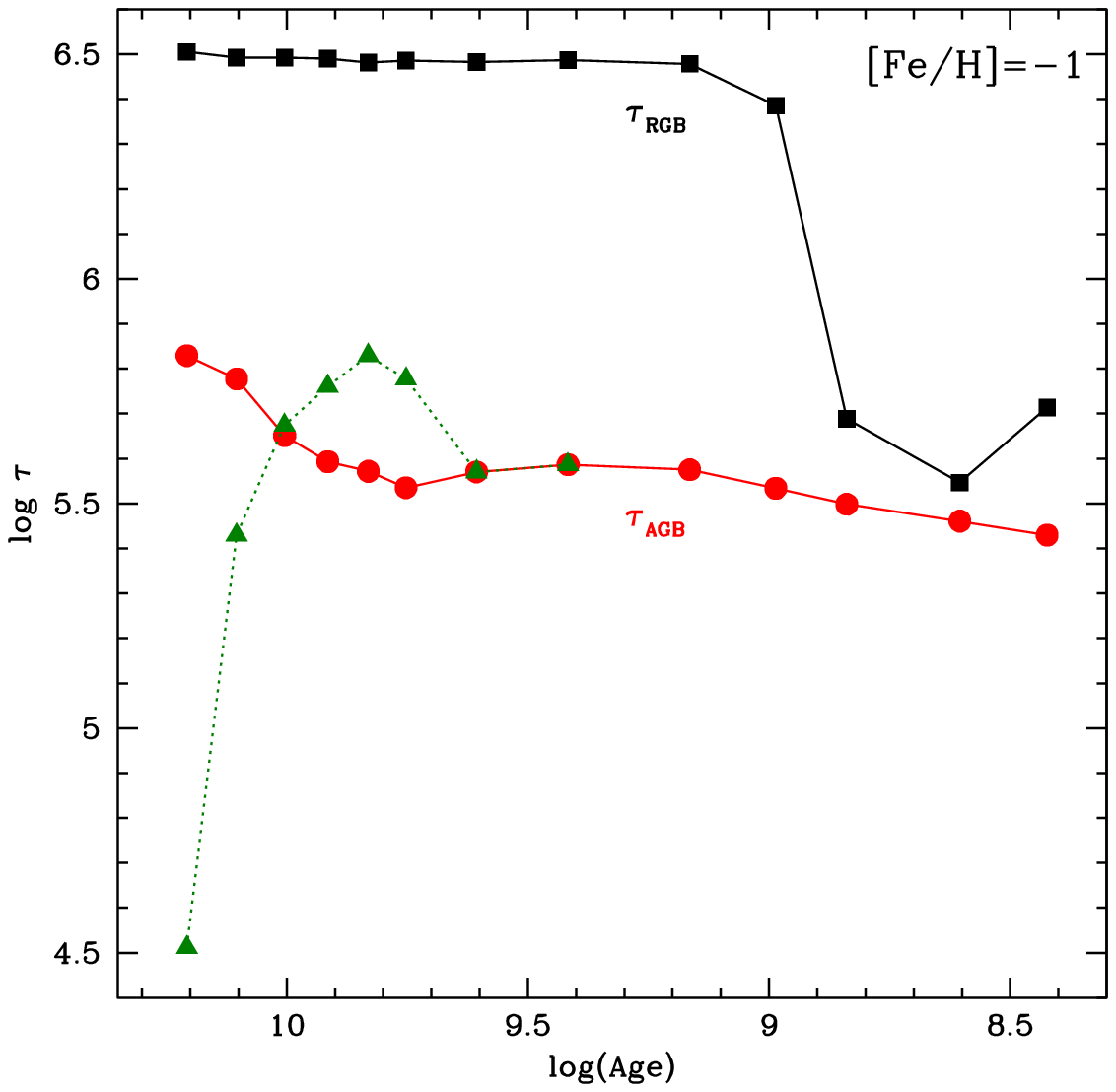}}
\end{minipage}
\vskip-60pt
\caption{Duration of the phase during which the stars evolve into
the RGB (black squares and lines) and the AGB (red dots and lines) boxes 
introduced by H23, as a 
function of the initial mass (left panel) and the formation epoch
(right). Dotted lines and green triangles refer to the results obtained
by assuming $\rm 0.2~M_{\odot}$ mass loss during the
RGB evolution of $\rm M<1.5~M_{\odot}$ stars.}
\label{ftimes}
\end{figure*}

To have a more general view of the behaviour of stars of different masses, 
we show in the lower panels of Fig.~\ref{f0820} the time evolution of 
$\rm F814W$ for $\rm 1.5~M_{\odot}$ and $\rm 2.5~M_{\odot}$ stars. For 
the $\rm 1.5~M_{\odot}$ star, which undergoes the helium flash, both the RGB 
phase preceding the ignition of core helium burning and the AGB phase are shown, 
whereas in the $\rm 2.5~M_{\odot}$ case the plot is limited to the early-AGB 
and the TP-AGB evolution.
In the $\rm 0.8~M_{\odot}$ case, the $\rm F814W$ flux increases as a
consequence of the gradual rise in the stellar luminosity, due
to the growth of the core mass, until the final
phases. Higher mass stars evolve differently,
as the formation of dust in the circumstellar envelope favours the
gradual shift of the SED to the IR, therefore the $\rm F814W$ flux
eventually decreases as the bolometric luminosity continues to increase. We note
in Fig.~\ref{f0820} that the whole TP-AGB evolution of $\rm M \leq 2~M_{\odot}$ 
stars takes place within the AGB box chosen by H23, whereas the stars
descending from higher mass progenitors, which evolve on time scales
below half a Gyr, reach brighter $\rm F814W$ fluxes. Therefore, a taller 
box (in magnitude) would give access to more massive AGB stars, giving 
information about more recent SFH.

\subsection{The times in the RGB and AGB boxes: the trend with the stellar mass}
To have an overall view of the duration of the different evolutionary
phases, we show in Fig.~\ref{ftimes} the times spent within the RGB 
($\rm \tau_{RGB}$) and the AGB ($\rm \tau_{AGB}$) boxes given in H23,
as a function of the initial mass (left panel) and of the age (right)
of the stars for the different model stars considered here.

$\rm \tau_{RGB}$ is of the order of 3 Myr for all stars of mass below 
$\rm 2~M_{\odot}$; this is a consequence of the onset of electron degeneracy, 
which makes the main features of the last part of the RGB evolution independent 
of the stellar mass.  
According to the earlier discussion regarding the evolution 
of the $\rm 0.8~M_{\odot}$ model star, the $\rm \tau_{RGB}$ of low-mass stars is further 
increased by $\sim 20\%$ by the second crossing of the evolutionary track 
with the RGB box, taking place during the early-AGB evolution.
A similar behaviour is seen in the $\rm 1.5~M_{\odot}$ model star depicted in 
the bottom, left panel of Fig.~\ref{f0820}. The time spent within the RGB box 
is about a factor 5-6 shorter for 
$\rm M \geq 2~M_{\odot}$ stars, because the
stars not undergoing the helium flash enter the RGB box during 
the early-AGB phase, which is significantly shorter than the RGB ascending.

$\rm \tau_{AGB}$ is much shorter than $\rm \tau_{RGB}$, while its trend 
with mass is not monotonic. If we consider the case where no mass loss took place during the RGB phase (red points in Fig.~\ref{ftimes}), $\rm \tau_{AGB}$ initially decreases with the mass of the star, because the lower $\rm M$, the slower the evolutionary time scales. At masses around $\rm 1~M_{\odot}$, this trend is reversed as the star experiences a higher number of TPs and spends its entire TP-AGB phase within the AGB box, as can be seen in the bottom-left panel of Fig.~\ref{f0820}. A further inversion of the $\rm \tau_{AGB}$-mass trend is found at $\rm \sim 1.5~M_{\odot}$, which is again a consequence of the shorter evolutionary time scales of higher mass stars. 

However, it is well established that stars lose mass during the RGB phase \citep{reimers77}, and
thus the hypothesis that the post-flash evolution starts with the same mass
with which they formed is unreasonable, particularly in stars of mass below
the solar mass. The occurrence of mass loss alters the scenario proposed above,
as will be described below.

\subsection{The role of mass loss during the RGB phase}
\label{mloss}
To investigate how mass loss during the RGB phase alters the 
expected number counts within the RGB and AGB boxes, we study how $\rm \tau_{RGB}$
and $\rm \tau_{AGB}$ change when the mass of the star decreases during the RGB phase. 
A preliminary idea on the effects of mass loss can
be obtained by looking at the top, left panel of Fig.~\ref{f0820}, where the time
evolution of the F814W magnitude of a $\rm 0.8~M_{\odot}$ model star calculated with
no RGB mass loss is compared with the results obtained
by assuming a total mass loss of $\rm \delta m_{RGB}=0.1~M_{\odot}$ 
and $\rm 0.2~M_{\odot}$. Mass loss does not affect $\rm \tau_{RGB}$, since 
the evolution of stars that develop a degenerate
core is primarily driven by the core mass.

On the other hand, $\rm \tau_{AGB}$ is affected by $\rm \delta m_{RGB}$:
the RGB mass loss leads to
a decrease in the mass with which the stars begin their AGB lifecycle, resulting in a smaller number of TPs, 
and hence a shorter overall duration of the AGB phase. This is evident in the case
of the $\rm 0.8~M_{\odot}$ model star,
where the number of TPs experienced is 6, 4, 3 for $\rm \delta m_{RGB}=0$, 
$\rm 0.1~M_{\odot}$, and $\rm 0.2~M_{\odot}$, respectively. Hence the occurrence 
of mass loss during the RGB phase of low-mass stars leads to lower 
values of $\rm N_{AGB}/N_{RGB}$.

To better understand the effects of the RGB mass loss 
we compare in  Fig.~\ref{ftimes} the results obtained by assuming 
$\rm \delta m_{RGB}=0$ and $\rm \delta m_{RGB}=0.2~M_{\odot}$. The 
differences are particularly relevant in the $\rm M \leq 0.8~M_{\odot}$ 
mass domain, owing to the notable 
decrease in the duration of the AGB phase of stars exposed
to significant mass loss during the RGB phase, which either do not enter the
TP-AGB phase, or experience only 1-2 TPs before the envelope is lost (compare the 
red and green lines in the top, left panel of Fig.~\ref{f0820}). 

The differences among the values of $\rm N_{AGB}/N_{RGB}$ obtained with 
different choices of $\rm \delta m_{RGB}$
are relevant both on quantitative and qualitative grounds. Still considering
the two cases shown in Fig.~\ref{ftimes}, we expect that in the no
RGB mass loss case $\rm N_{AGB}/N_{RGB}$ would decrease with time starting 
from the most remote epochs, while when $\rm \delta m_{RGB}=0.2~M_{\odot}$ is
considered $\rm N_{AGB}/N_{RGB}$ would increase until $\sim 6$ Gyr ago, then
decrease afterwards. Higher $\rm \delta m_{RGB}$'s would reflect into
a later occurrence of the maximum in $\rm N_{AGB}/N_{RGB}$.

\section{AGB/RGB ratio in the Local Group satellites}
\label{galaxies}
In the following, we concentrate on some selected galaxies of metallicity
$\rm [Fe/H]=-1$, for which the SFH was derived in previous studies.
We compare the $\rm N_{AGB}/N_{RGB}$ ratio 
obtained by means of a population synthesis approach with the
values estimated via number counts of the sources in the RGB and
AGB boxes considered. For each epoch, we assume that the mass distribution 
of the stars formed follows \cite{kroupa01}'s initial mass function.
We consider three different groups, selected
on the basis of the SFH experienced.

\subsection{The ''intermediate cases'' of Andromeda I, II}
The SFH of Andromeda I and Andromeda II is characterised by a first, 
intense episode, concluded around 10 Gyr ago, and further significant 
star formation in more recent times \citep{Weisz14}.

\begin{figure*}
\begin{minipage}{0.48\textwidth}
\resizebox{1.\hsize}{!}{\includegraphics{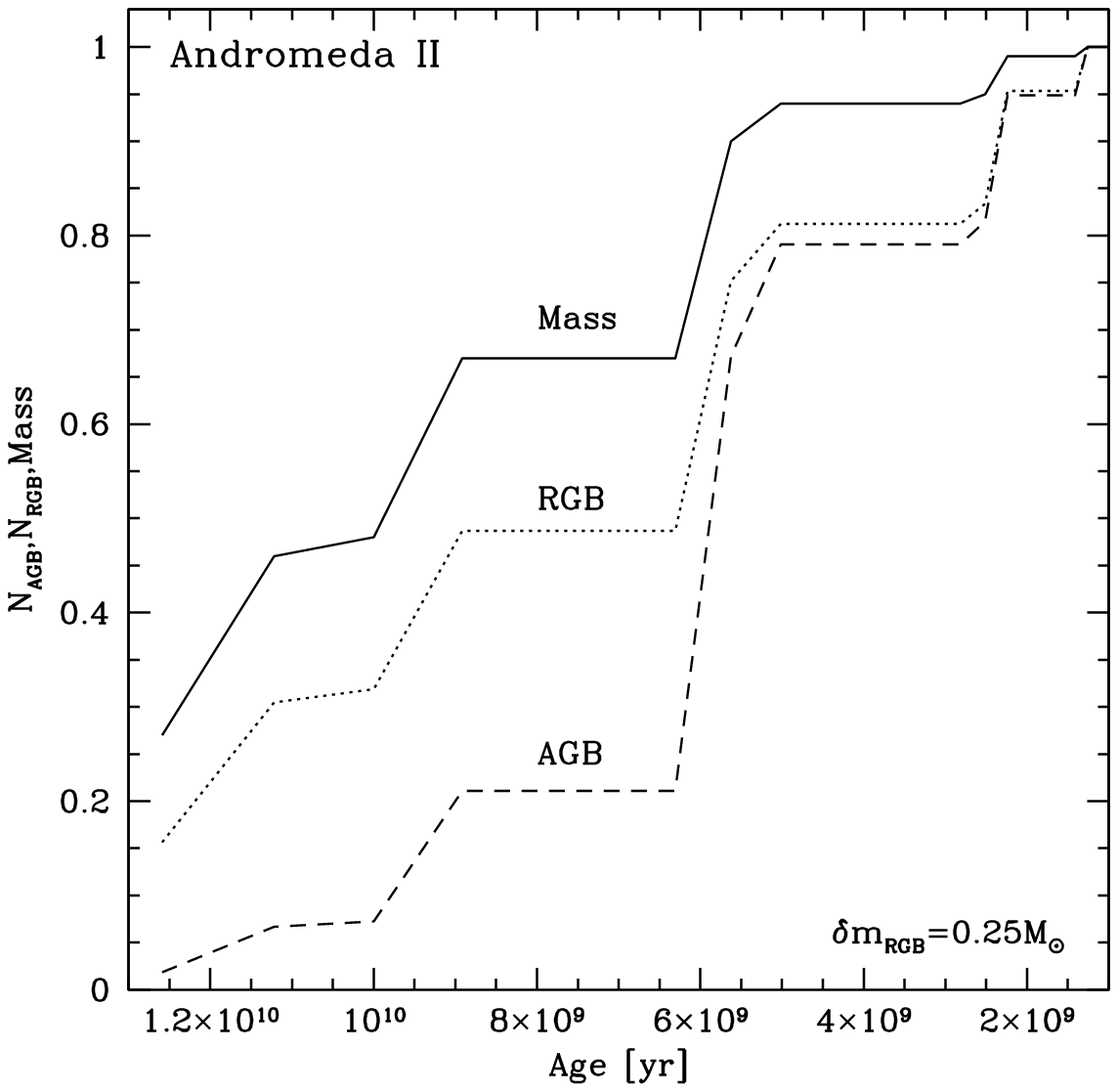}}
\end{minipage}
\begin{minipage}{0.48\textwidth}
\resizebox{1.\hsize}{!}{\includegraphics{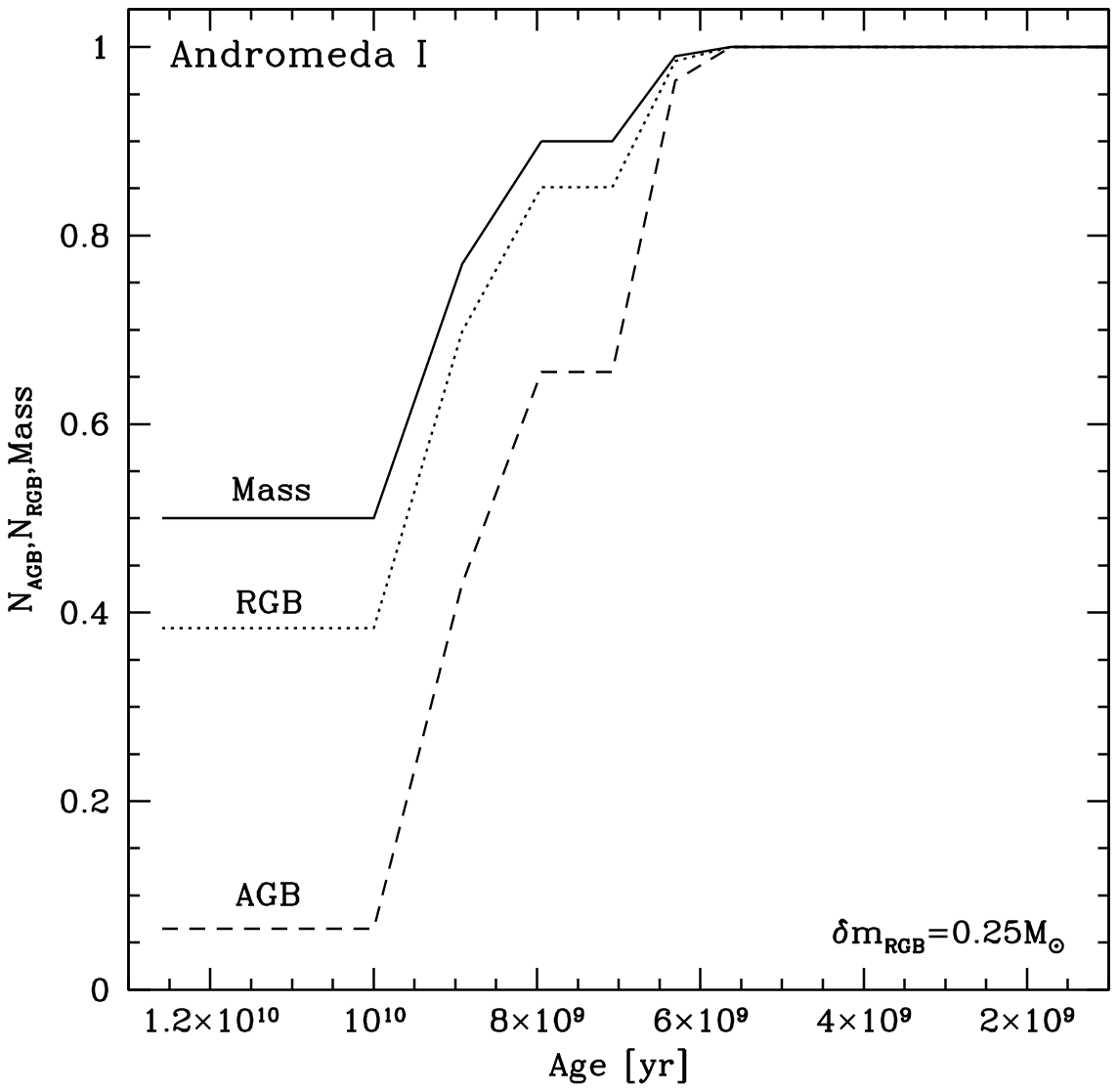}}
\end{minipage}
\vskip-100pt
\begin{minipage}{0.48\textwidth}
\resizebox{1.\hsize}{!}{\includegraphics{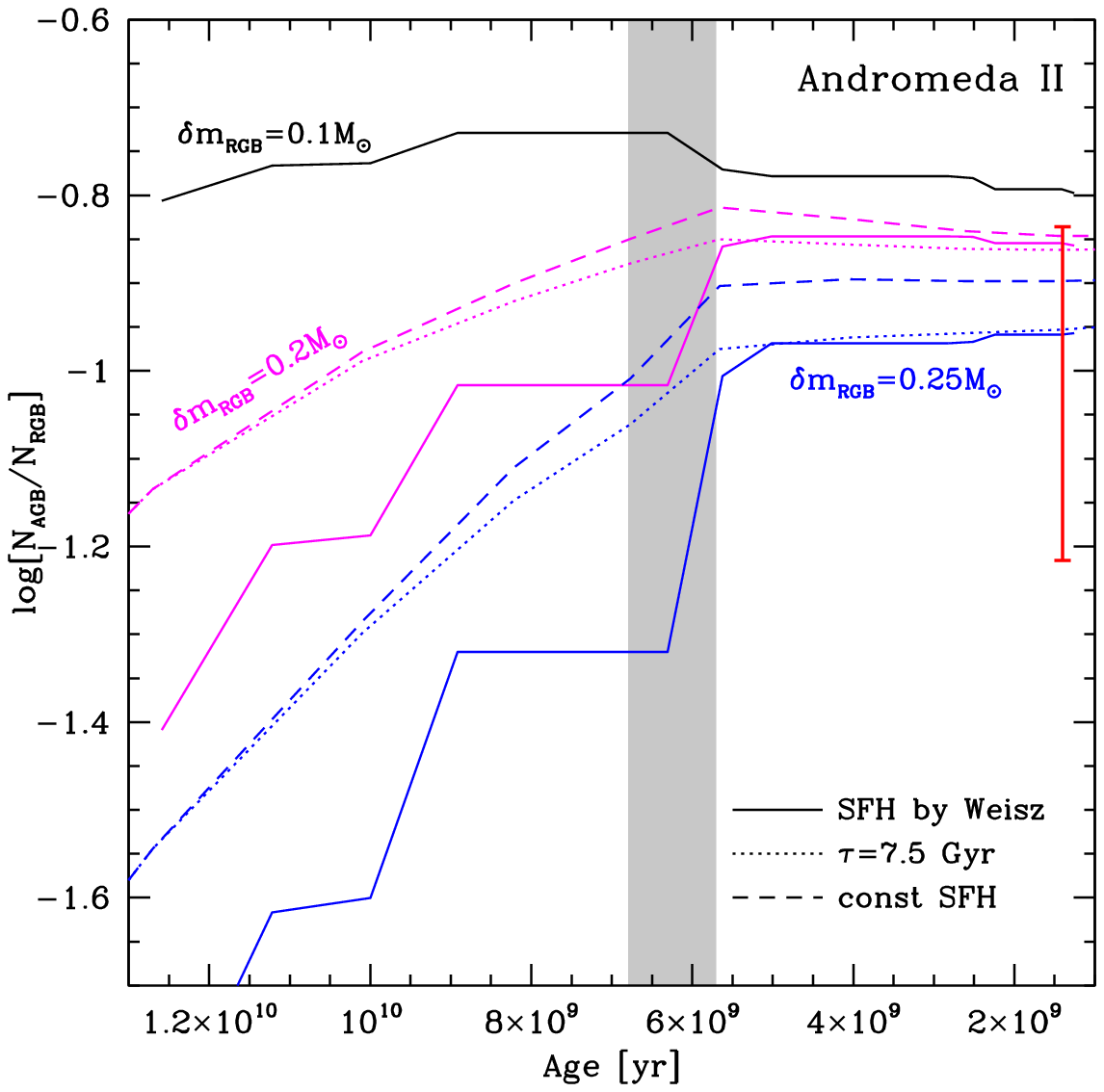}}
\end{minipage}
\begin{minipage}{0.48\textwidth}
\resizebox{1.\hsize}{!}{\includegraphics{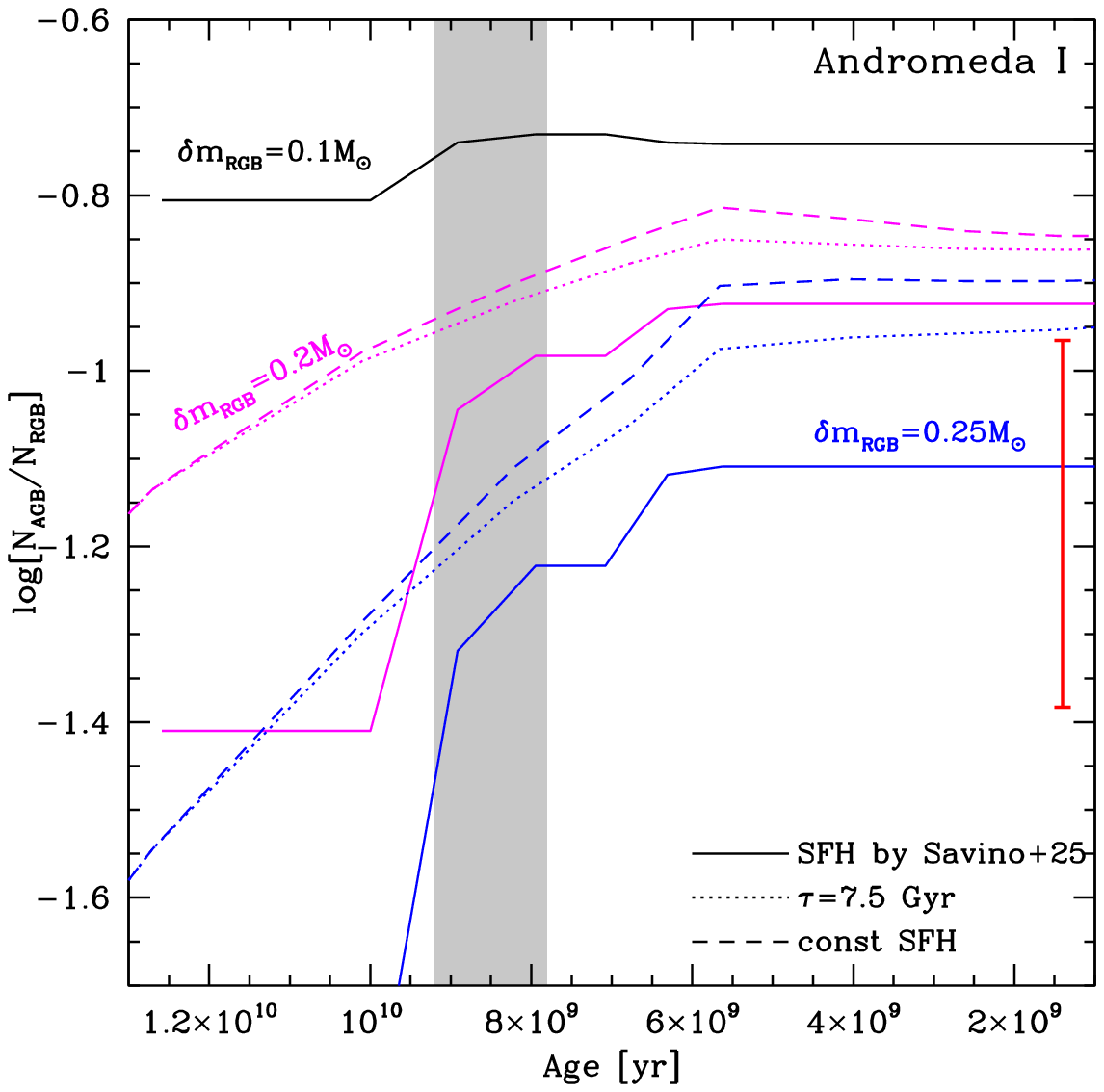}}
\end{minipage}
\vskip-60pt
\caption{The variation of the relative mass of the galaxy
(solid line), of the fraction of AGB (dashed) and RGB (dotted) stars formed,
for the $\rm \delta m (RGB)= 0.25~M_{\odot}$ case and the SFH by \citet{Weisz14}
and \citet{savino25} for the galaxies Andromeda II (top, left panel) and Andromeda I (top, right), 
respectively. The bottom panels show
the cumulative contribution of various star formation epochs 
to the present-day ratio of AGB to RGB stars of Andromeda II 
(bottom, left panel) and Andromeda I (bottom, right panel), based on different assumptions 
regarding the star formation history and the average mass lost by low-mass 
stars during the RGB phase. Solid lines indicate results found by adopting 
the SFH by \citet{Weisz14}, for Andromeda II, and \citet{savino25}, for 
Andromeda I; dashed tracks are based on the assumption of constant SFH; 
dotted lines refer to  results obtained when assuming that the SFH decays 
with a time scale of 7.5 Gyr. Colour coding indicates the mass lost by stars 
during the RGB evolution: the cases $\rm \delta m (RGB)= 0.1~M_{\odot}$ (black
lines), $\rm 0.2~M_{\odot}$ (magenta), $\rm 0.25~M_{\odot}$ (blue) 
are shown. The grey shaded region indicates the T90 epoch, the red 
line refers to the AGB to RGB ratio derived for Andromeda II and Andromeda I.
}
\label{fandII}
\end{figure*}

Andromeda II, as shown in the top, left panel of
Fig.~\ref{fandII}, experienced three main episodes of star formation: 
a) $\sim 67\%$ of the stellar mass formed 
earlier than 9 Gyr ago; b) $\sim 27\%$ formed 
between 5 and 6.5 Gyr ago; c) the remaining $\sim 5\%$ stars are younger 
than 2.5 Gyr. Considering the evolutionary time scales of model 
stars of different mass, we conclude that the stars formed during the
periods (a), (b) and (c) now evolving through the RGB and AGB phases
descend from progenitors of mass 
$\rm \sim 0.8~M_{\odot}$, $\rm 1-1.1~M_{\odot}$, and $\rm 1.5-2~M_{\odot}$, 
respectively. 

The bottom, left panel of Fig.~\ref{fandII} shows the contribution of various epochs
of the SFH to the present-day $\rm N_{AGB}/N_{RGB}$. The solid lines indicate 
results based on the SFH by \citet{Weisz14},  
and $\rm \delta m_{RGB} = 0.1~M_{\odot}$, $\rm 0.2~M_{\odot}$, $\rm 0.25~M_{\odot}$.
We note the large differences in the contribution of the early epochs 
to the final $\rm N_{AGB}/N_{RGB}$ obtained for the different $\rm \delta m_{RGB}$'s. 
This is because the majority of the stars of Andromeda II formed during this period,
and the $\rm \tau_{AGB}$ of low-mass stars is tightly connected 
to $\rm \delta m_{RGB}$ (see Fig.~\ref{ftimes}):
for the largest $\rm \rm \delta m_{RGB}$'s the overall AGB phase is skipped.

In the first important phase of star formation of the galaxy, corresponding
to the point a) above, the contribution 
of increasingly more massive stars which spend more and more time in the AGB
boxes (see Fig.~\ref{ftimes}) serves to increase the final $\rm N_{AGB}/N_{RGB}$ 
as a function of formation time. A further rise in $\rm N_{AGB}/N_{RGB}$ takes place during 
the second burst of star formation of Andromeda II, which took place around 6 Gyr ago. The individual relative fractions of the stars in the RGB and AGB
boxes are reported in the bottom, left panel of Fig.~\ref{fandII}, for the
$\rm \delta m_{RGB}=0.25~M_{\odot}$ case. On the other hand, in the $\rm \delta m_{RGB}=0.1~M_{\odot}$ case the cumulative $\rm N_{AGB}/N_{RGB}$ is not monotonic with formation time and even decreases, following the behaviour of $\rm \tau_{AGB}/\tau_{RGB}$, as shown in Fig.~\ref{ftimes}.

For what regards the stars nowadays populating the
RGB box, independently of the assumption regarding $\rm \delta m_{RGB}$, 
we find that half descend from $\rm \sim 0.8~M_{\odot}$ progenitors,
$\sim 35\%$ from $\rm \sim 1~M_{\odot}$ stars, $\sim 15\%$ are 
younger objects, of initial mass above solar.
These percentages approximately reflect the population
of the AGB box in the $\rm \delta m_{RGB}=0.1~M_{\odot}$ case. On the other
hand, when $\rm \delta m_{RGB}=0.2~M_{\odot}$ is assumed, 
the percentage of low-mass stars within the
AGB box drops to $\sim 35\%$, so that the dominant population would be the
progeny of $\rm 1-1.1~M_{\odot}$ stars, accounting for $\sim 50\%$. 
In the $\rm \delta m_{RGB}=0.25~M_{\odot}$ case the oldest objects account 
for only $\sim 20\%$ of the AGB box, which would be mainly populated by $\sim 60\%$ of the progeny of $\rm \sim 1~M_{\odot}$ stars, the remaining $\sim 20\%$ being younger stars. 

The differences in the expected star counts within the AGB box
reflect into a strong sensitivity of the current $\rm N_{AGB}/N_{RGB}$ 
on $\rm \delta m_{RGB}$: we find that $\rm N_{AGB}/N_{RGB}=0.11, 0.14, 0.16$
when $\rm \delta m_{RGB}=0.25, 0.2, 0.1~M_{\odot}$, respectively.
The result by H23, $\rm N_{AGB}/N_{RGB}=0.071 \pm 0.037$, also shown on the
right side of the bottom, left panel of Fig.~\ref{fandII}, is consistent
with $\rm \delta m_{RGB} = 0.2-0.25~M_{\odot}$. 

We also discuss the cases that: a) the SFR of Andromeda II has been 
constant at all epochs until now; b) the SFR decayed 
with a time scale of 7.5 Gyr. The corresponding results are shown with dashed 
and dotted lines, respectively, in Fig.~\ref{fandII}. We applied these 
hypothesis to the $\rm \delta m_{RGB} = 0.2-0.25~M_{\odot}$ cases only, as
the results shown in Fig.~\ref{fandII} seem to rule out the 
$\rm \delta m_{RGB} = 0.1~M_{\odot}$ assumption. 
The $\rm N_{AGB}/N_{RGB}$'s obtained with the assumptions a) and b) above
are generally higher than those obtained with the SFH given in
\citet{Weisz14}, because in the latter case the mass distribution is more
peaked towards the lower mass domain.

The SFH of Andromeda I (see the top, right panel of Fig.~\ref{fandII}) 
is characterised by a more continuous trend with respect
to Andromeda II, with a gradual growth until $\sim 6$ Gyr ago, followed
by a $\sim 2.5$ Gyr break, and by a final episode that continued until
$\sim 1.5$ Gyr ago. Around $92\%$ of the stars in the galaxy have ages
above $\sim 6$ Gyr. The timing of the star formation reflects into
the evolved population presently evolving in the RGB box, $\sim 80\%$  of 
which is composed of stars descending from sub-solar mass progenitors,
while the remaining $\sim 20\%$ is the progeny of $\rm 1.2-1.5~M_{\odot}$ stars. 

The bottom, right panel of Fig.~\ref{fandII} shows the contribution of various epochs
of the star formation history to the present-day $\rm N_{AGB}/N_{RGB}$. 
As in the case of Andromeda II, the contribution of 
the early epochs to $\rm N_{AGB}/N_{RGB}$ is sensitive to the choice of 
$\rm \delta m_{RGB}$. In the $\rm \delta m_{RGB}=0.1~M_{\odot}$ case, the 
trend is approximately flat, since all low-mass stars evolve into the AGB 
box, thus  $\rm N_{AGB}/N_{RGB}$ keeps around 0.15, consistently with the 
results shown in Fig.~\ref{ftimes}. 
In the $\rm \delta m_{RGB} \geq 0.2~M_{\odot}$ cases the 
lowest mass stars barely reach the AGB box: this is the
reason for the severe drop in the contribution of the earlier epochs to 
$\rm N_{AGB}/N_{RGB}$ values, seen in the bottom, right panel of 
Fig.~\ref{fandII}.

A further significant feature in the run of $\rm N_{AGB}/N_{RGB}$
is the change in the slope of the $\rm N_{AGB}/N_{RGB}$ vs. time
at $\sim 6$ Gyr, when the aforementioned break in the star 
formation occurred. No significant further increase in 
$\rm N_{AGB}/N_{RGB}$ takes place during the final epochs, thus the final
ratio is mostly determined by the value attained before $\sim 6$ Gyr
ago. 

The comparison between the synthetic and the observed values
of $\rm N_{AGB}/N_{RGB}$ outlines consistency only in the 
$\rm \delta m_{RGB}=0.25~M_{\odot}$ case. In this case results
from synthetic modelling indicate that $75\%$ of the stars currently
in the AGB box descend from progenitors of mass below solar, 
and that most of this population is the progeny of $\rm 0.9-1~M_{\odot}$
stars. The RGB box has a wider mass distribution, with a significant 
presence ($32\%$) of $\rm 0.8-0.9~M_{\odot}$ stars.

The analysis of Andromeda I and Andromeda II indicates that
the observed $\rm N_{AGB}/N_{RGB}$ values can be reproduced only
if we assume that low mass stars lose at least $\rm 0.2~M_{\odot}$
while evolving through the RGB. Interestingly, given the
particular evolution of the star formation rate of these galaxies,
the results obtained are not substantially different from
those based on the assumption of a star formation
rate decaying with a 7.5 Gyr time-scale.

\subsection{The oldest populations: the Sculptor and NGC 185 cases}
In Sculptor and NGC 185 most of the star formation took place before
10 Gyr ago. Therefore, the mass distribution is 
strongly peaked towards the lowest masses.

\begin{figure*}
\begin{minipage}{0.48\textwidth}
\resizebox{1.\hsize}{!}{\includegraphics{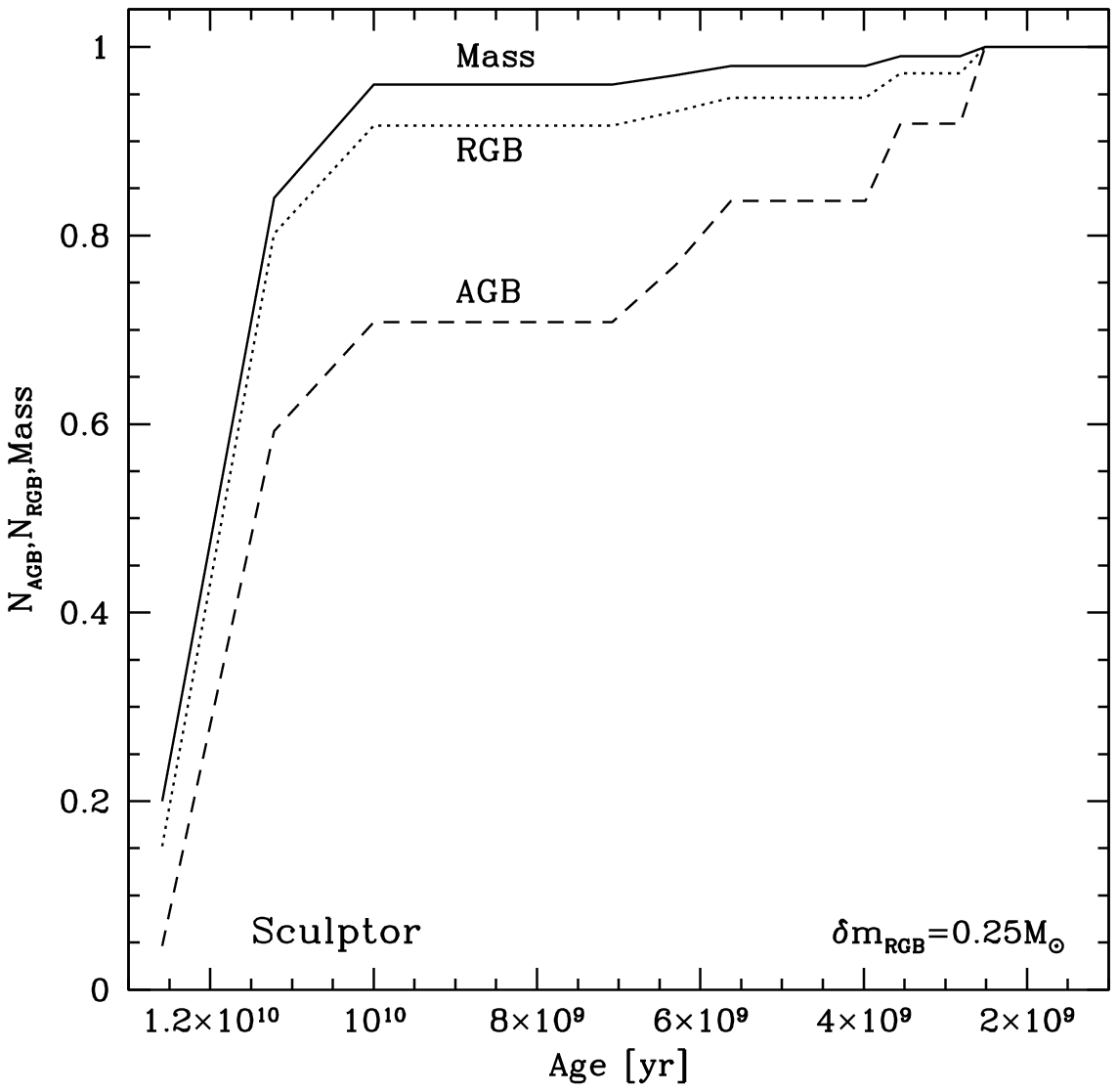}}
\end{minipage}
\begin{minipage}{0.48\textwidth}
\resizebox{1.\hsize}{!}{\includegraphics{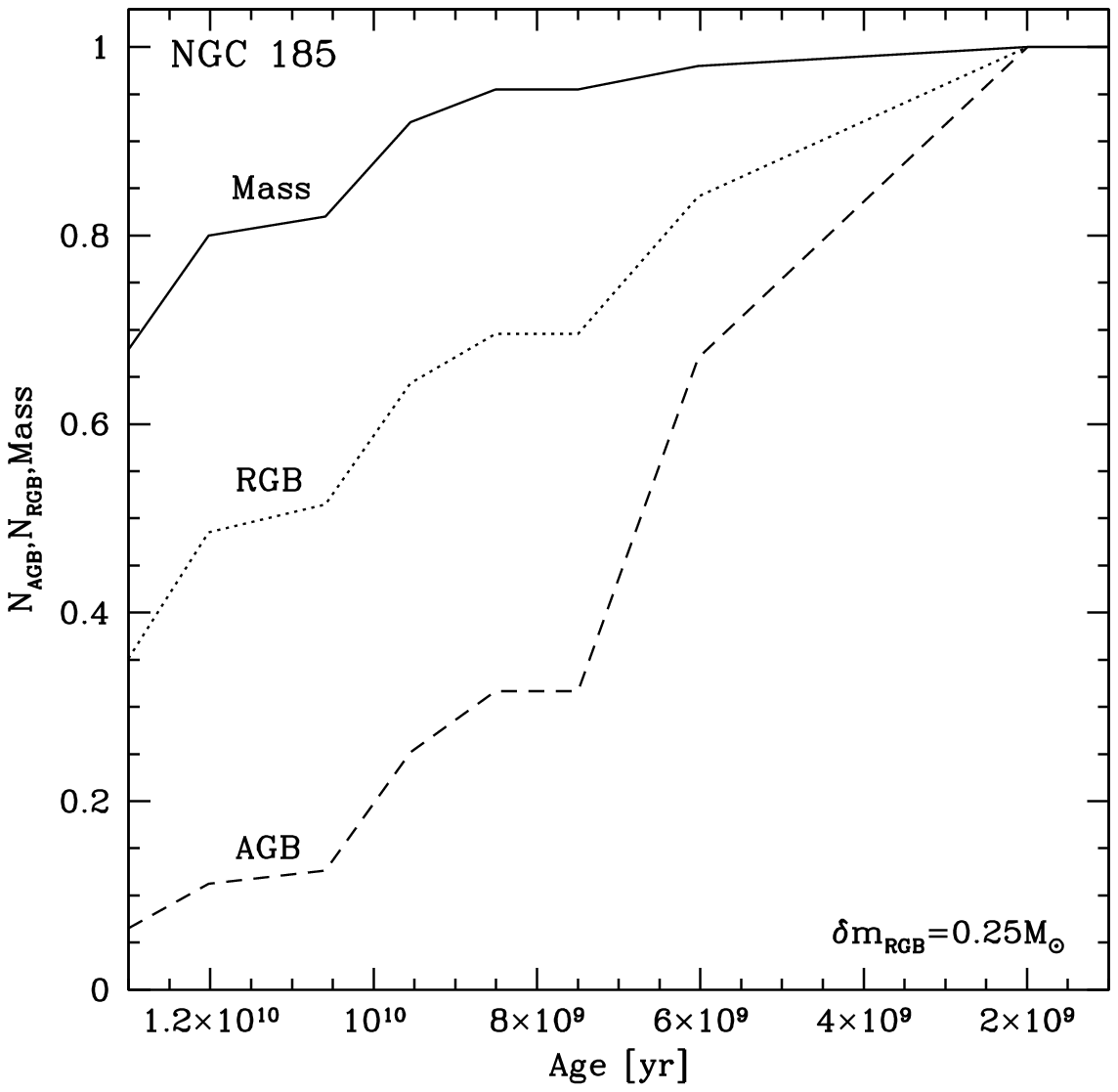}}
\end{minipage}
\vskip-100pt
\begin{minipage}{0.48\textwidth}
\resizebox{1.\hsize}{!}{\includegraphics{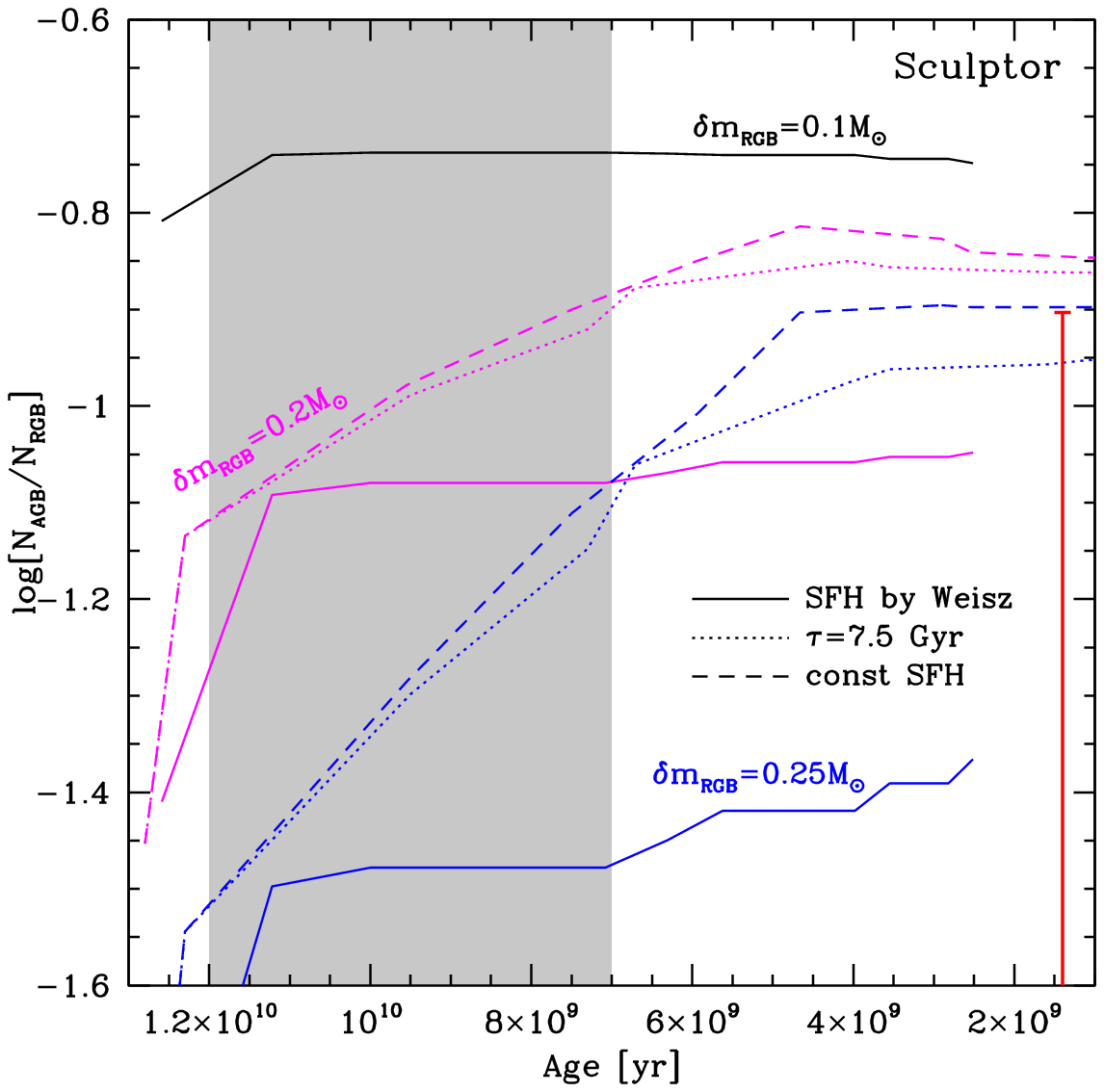}}
\end{minipage}
\begin{minipage}{0.48\textwidth}
\resizebox{1.\hsize}{!}{\includegraphics{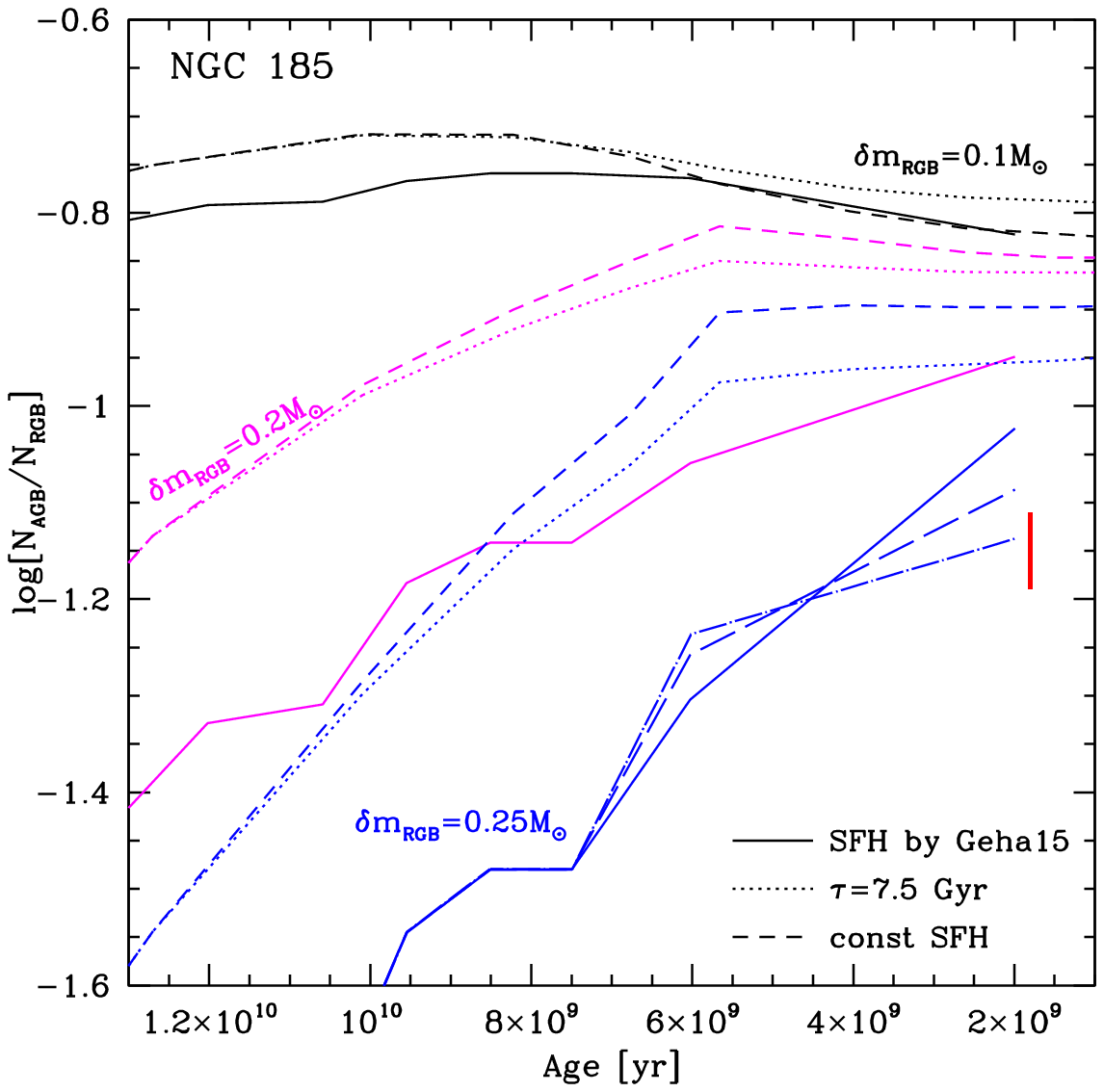}}
\end{minipage}
\vskip-60pt
\caption{The variation of the relative fractions of the
mass of the galaxy and of the stars in the RGB and AGB boxes for
the $\rm \delta m_{RGB}= 0.25~M_{\odot}$ case, based on the SFH by 
\citet{Weisz14} for Sculptor (top, left panel) and NGC185 (top, right).
The bottom panels show the cumulative contribution of various star formation epochs 
to the present-day ratio of AGB to RGB stars of Sculptor
(bottom, left panel) and NGC185 (bottom, right), based on different assumptions 
regarding the star formation history and the average mass lost by low-mass 
stars during the RGB phase. Solid lines indicate results found by adopting 
the SFH by \citet{Weisz14} for Sculptor and for NGC 185. The colour coding, 
as also the dashed and dotted lines, have the same meaning as in Fig.~\ref{fandII}.  
}
\label{fscul}
\end{figure*}

In regard to Sculptor, as shown in the top, left panel of
Fig.~\ref{fscul}, $96\%$ of the stellar mass formed earlier
than 10 Gyr ago, thus $\sim 90\%$ of the current evolved stellar population is
composed by the progeny of $\rm \sim 0.8~M_{\odot}$ stars. Star formation 
stopped $\sim 2.5$ Gyr ago, thus the present-day RGB and AGB populations
descend from progenitors of mass below $\rm 1.5~M_{\odot}$.

The contribution of various epochs to the present-day $\rm N_{AGB}/N_{RGB}$ 
of Sculptor obtained on the basis of different assumptions is reported in 
the bottom, left panel of Fig.~\ref{fscul}. The evolution of the relative
fractions of the stars in the RGB and AGB boxes, for the 
$\rm \delta m_{RGB}= 0.25~M_{\odot}$ case only, are shown in the
top, left panel of Fig.~\ref{fscul}.

In the $\rm \delta m_{RGB}= 0.1~M_{\odot}$ case, the initial mass distribution of the stars
contained in the RGB and AGB boxes is similar and peaked around $\rm 0.8~M_{\odot}$.
As a consequence, $\rm N_{AGB}/N_{RGB}$ attains 
values $\sim 0.15$, which reflect the ratio of the relative duration of the AGB and RGB phases 
typical of $\rm 0.8~M_{\odot}$ stars, as shown in 
Fig.~\ref{ftimes}. No meaningful changes in $\rm N_{AGB}/N_{RGB}$ are found during the more
recent epochs, given the dominant contribution from $\sim 0.8~M_{\odot}$ in the
two boxes considered.

In the $\rm \delta m_{RGB}= 0.2~M_{\odot}$ and $\rm 0.25~M_{\odot}$ cases, the
stars of initial mass $\rm \sim 0.8~M_{\odot}$ evolve within the AGB box
for a shorter time than for $\rm \delta m_{RGB}= 0.1~M_{\odot}$:
therefore, the percentage of $\rm \sim 0.8~M_{\odot}$ stars in the AGB box decreases, 
down to $72\%$ and $63\%$, respectively. The final $\rm N_{AGB}/N_{RGB}$
ratios are consequently lower than for $\rm \delta m_{RGB}= 0.1~M_{\odot}$,
with $\rm N_{AGB}/N_{RGB}$=0.09, for $\rm \delta m_{RGB}= 0.2~M_{\odot}$,
and $\rm N_{AGB}/N_{RGB}$=0.04, for $\rm \delta m_{RGB}= 0.25~M_{\odot}$.
The comparison with the $\rm N_{AGB}/N_{RGB}$ estimated from number counts
rules out the $\rm \delta m_{RGB}= 0.1~M_{\odot}$ possibility, while both the
$\rm \delta m_{RGB}= 0.2~M_{\odot}$ and $\rm 0.25~M_{\odot}$ cases are
consistent with the observations.

For what concerns NGC 185, the rise of the SFR with 
time during the early 
epochs is slower than for Sculptor; this can be decuced
by comparing the evolution of the fractional mass
of the galaxies in the top, left and top, right panels
of Fig.~\ref{fscul}. The first major episode of star formation,
during which $95\%$ of the stellar mass formed,
ended around 8 Gyr ago. We see in the bottom, right panel
of Fig.~\ref{fscul} that $\rm N_{AGB}/N_{RGB}$ increases
until that epoch, reaching values similar
to those of Sculptor, before the final rise, related
to a recent star formation episode \citep{geha15}. During this
episode, a significant 
number of $\rm M>1~M_{\odot}$ stars formed:
indeed the presence of these stars increases the number of stars
nowadays evolving within the AGB box, and consequently $\rm N_{AGB}/N_{RGB}$.
We explored the sensitivity of the final $\rm N_{AGB}/N_{RGB}$
to the duration of the latter period of recent star formation,
which reflects into the overall stellar mass formed recently: we found 
that when the fraction of the stars formed in the recent epochs decreases 
from $2 \%$ to $0.5 \%$, $\rm N_{AGB}/N_{RGB}$
decreases by $\sim 0.15$ dex.

The errors associated with the estimate of $\rm N_{AGB}/N_{RGB}$ of
NGC 185 are smaller than for Sculptor, which allows a tighter constraint
on the RGB mass loss: $\rm \delta m_{RGB}= 0.25~M_{\odot}$
is the only choice consistent with the observations, in agreement 
with the conclusions drawn during the analysis of Andromeda I.
The corresponding evolution of the relative fractions of AGB and
RGB stars can be seen in the top, right panel of Fig.~\ref{fscul}.

In the cases of Sculptor and NGC 185, unlike Andromeda I and II, 
the final $\rm N_{AGB}/N_{RGB}$ obtained by
assuming either constant or exponentially decaying SFR  are far in 
excess of those based on the detailed SFH
available in the literature. This is due to the dominant
presence of low-mass stars, which is poorly reproduced by the constant or
the slowly decaying star formation rate hypothesis.

\subsection{The galaxies with recent star formation}
\begin{figure*}
\begin{minipage}{0.48\textwidth}
\resizebox{1.\hsize}{!}{\includegraphics{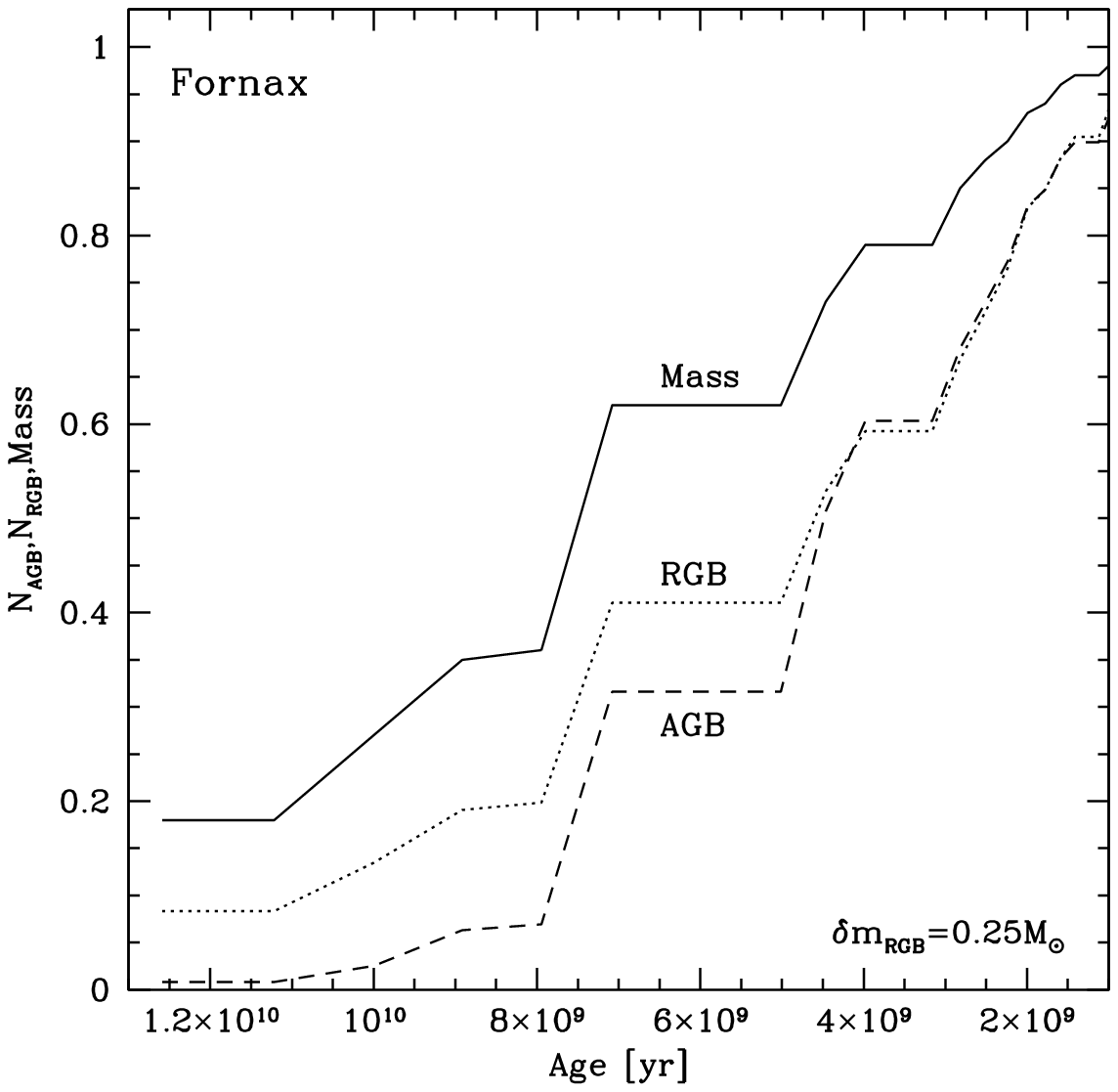}}
\end{minipage}
\begin{minipage}{0.48\textwidth}
\resizebox{1.\hsize}{!}{\includegraphics{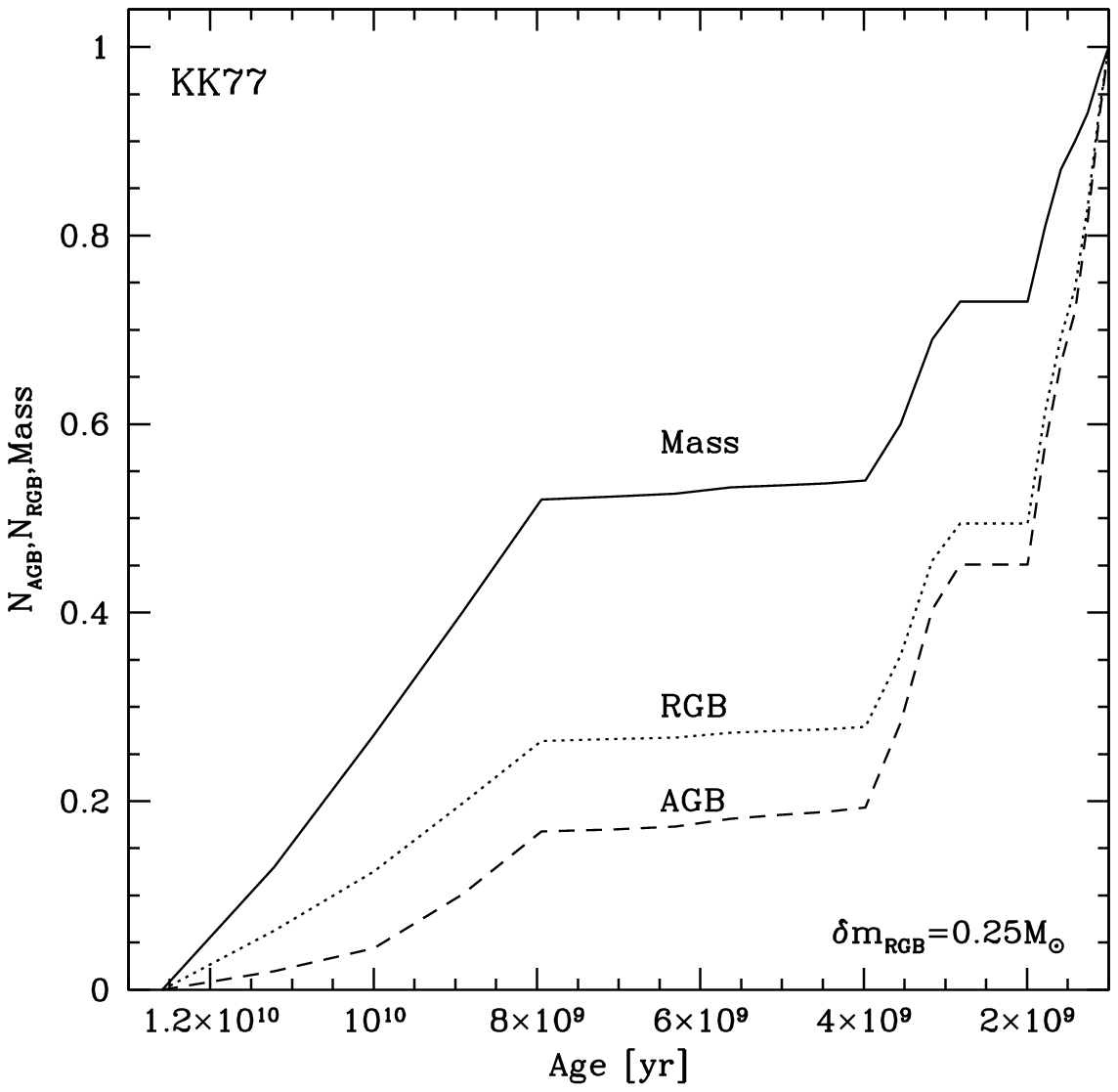}}
\end{minipage}
\vskip-100pt
\begin{minipage}{0.48\textwidth}
\resizebox{1.\hsize}{!}{\includegraphics{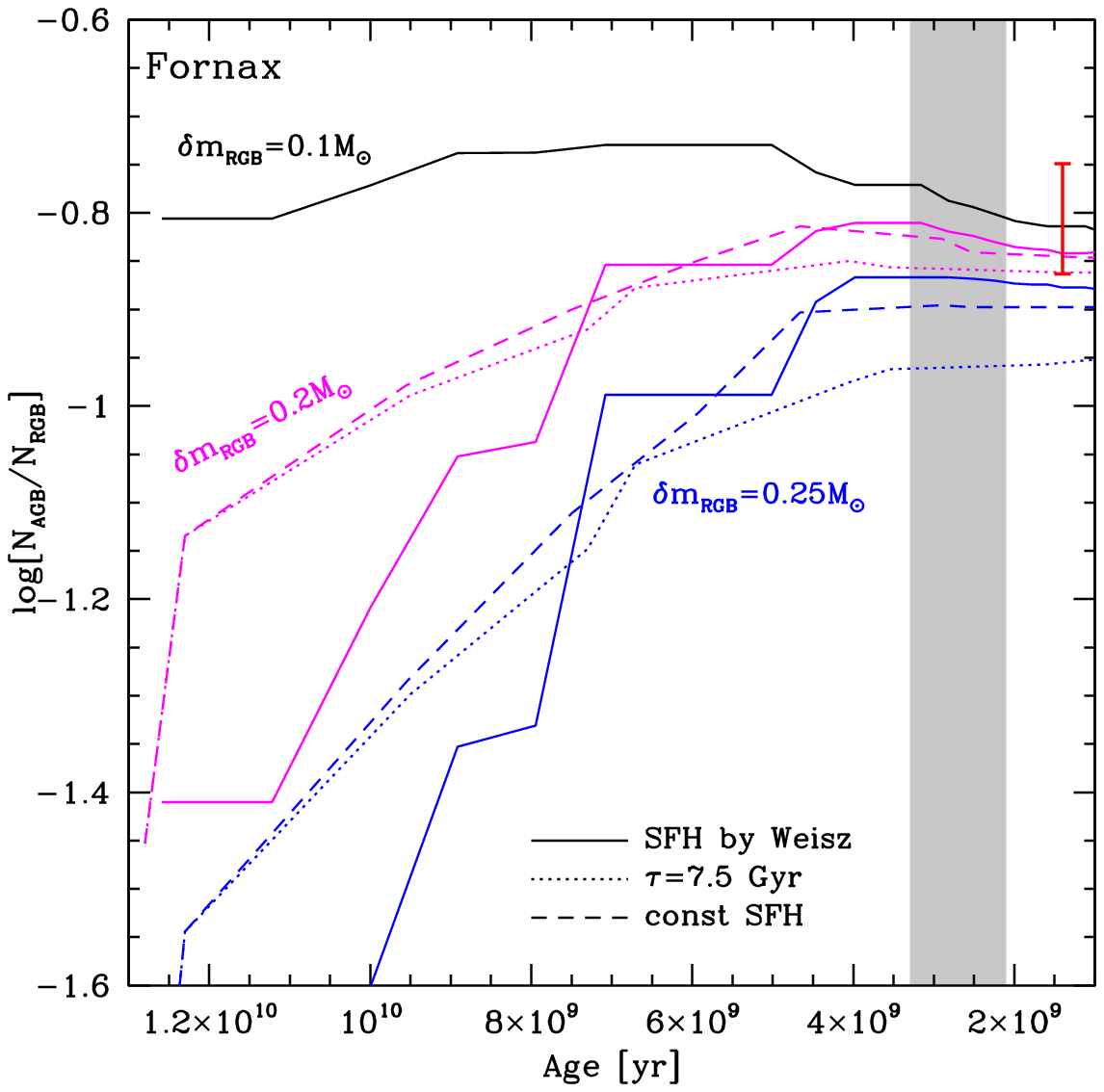}}
\end{minipage}
\begin{minipage}{0.48\textwidth}
\resizebox{1.\hsize}{!}{\includegraphics{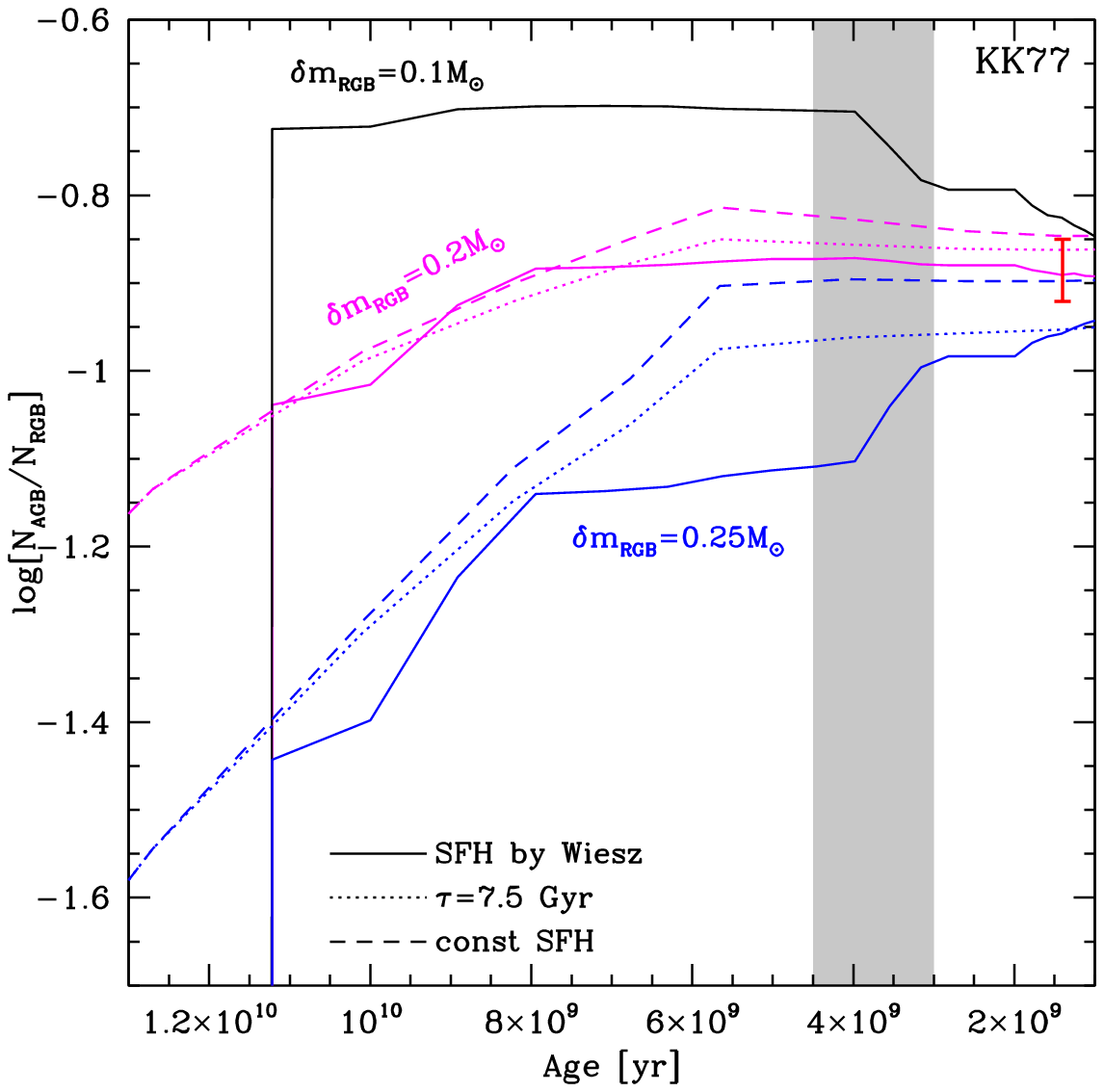}}
\end{minipage}
\vskip-60pt
\caption{The variation of the relative fractions of the
mass of the galaxy and of the stars in the RGB and AGB boxes for
the $\rm \delta m_{RGB}= 0.25~M_{\odot}$ case, based on the SFH by 
\citet{Weisz14} for Fornax (top, left panel) and KK77 (top, right).
The bottom panels show the cumulative contribution of various star formation epochs 
to the present-day ratio of AGB to RGB stars of Fornax 
(bottom, left panel) and KK77 (bottom, right). Solid lines
refer to results based on the SFH from \citet{Weisz14}. The meaning of
the dashed and dotted lines and of the grey region, and the colour coding,
is the same as in Fig.~\ref{fandII} and \ref{fscul}.
}
\label{ffornax}
\end{figure*}

In the galaxies such as Fornax and KK77, star formation continued 
until less than 1 Gyr ago. This can be deduced by the results 
reported in the top panels of Fig.~\ref{ffornax}, where the evolution
of the total mass (solid line) and of the variation of the stars in 
the RGB and AGB boxes (dotted and dashed line, respectively) are
shown. The bottom panels of 
the same figure show the cumulative contribution of various epochs to the 
final $\rm N_{AGB}/N_{RGB}$. 

We stress here the significant uncertainty associated with the SFH of KK77 
(observed as part of the ANGST survey, \citet{Dalcanton2009}), which was estimated from data that reach a little 
beyond the red clump, and does not include the main sequence turn-off stars.
In the bottom, right panel of Fig.~\ref{ffornax}, we refer to the results obtained 
by assuming the lower limit of the SFH derived by \citet{Weisz14} for this galaxy.

The continuity with which star formation occurred in these galaxies is confirmed 
by the similarity between the results based on the SFH published by
\citet{Weisz14} and those obtained by assuming a constant
SFH during the various epochs.

The study of Fornax and KK77 shows that for galaxies characterised by recent 
star formation the final $\rm N_{AGB}/N_{RGB}$ is scarcely dependent on the choice 
regarding the RGB mass loss. Indeed we see in the bottom panels of Fig.~\ref{ffornax}
that the results based on the various choices for
$\rm \delta m_{RGB}$ are all consistent with the measured
values of $\rm N_{AGB}/N_{RGB}$. The reason for this behaviour is that the majority 
of the stars nowadays evolving in the RGB and AGB boxes descend from 
progenitors whose mass is in the $\rm 0.9-1.2~M_{\odot}$ range:
for these stars, the time spent within the AGB box is not
extremely sensitive to $\rm \delta m_{RGB}$, as it is for
their lower mass counterparts.

In the case of Fornax, $\sim 80\%$ of the stars in the AGB box
and $70\%$ of those in the RGB box descend from $\rm 0.9-1.2~M_{\odot}$ 
progenitors. The final $\rm N_{AGB}/N_{RGB}$, of the order of
0.15, is in nice agreement with that derived from number counts,
within the error bar.

\section{Understanding the $\rm N_{AGB}/N_{RGB}$ ratios of
metal poor galaxies}
The results presented in the previous section outlined that the present-day
$\rm N_{AGB}/N_{RGB}$ of the galaxies in which most of the star formation 
occurred during the early epochs is largely sensitive to 
the amount of mass lost by stars during their RGB evolution:
consistency between the results from synthetic modelling and the measured
number counts demands $\rm \delta m_{RGB} > 0.2~M_{\odot}$. 
This value is consistent with the study by \citet{tailo20}, 
who estimated an average mass loss slightly above $\rm 0.2~M_{\odot}$ for 
RGB stars belonging to Galactic globular clusters of metallicity $\rm [Fe/H]=-1$.
The role of $\rm \delta m_{RGB}$ is less relevant for the
galaxies characterised by recent star
formation and $\rm T_{90} < 4$ Gyr: in these cases a $\rm 0.2~M_{\odot}$ change 
in $\rm \delta m_{RGB}$ would reflect into differences in the estimated 
$\rm N_{AGB}/N_{RGB}$ below 0.05 dex.

\begin{figure*}
\begin{minipage}{0.48\textwidth}
\resizebox{1.\hsize}{!}{\includegraphics{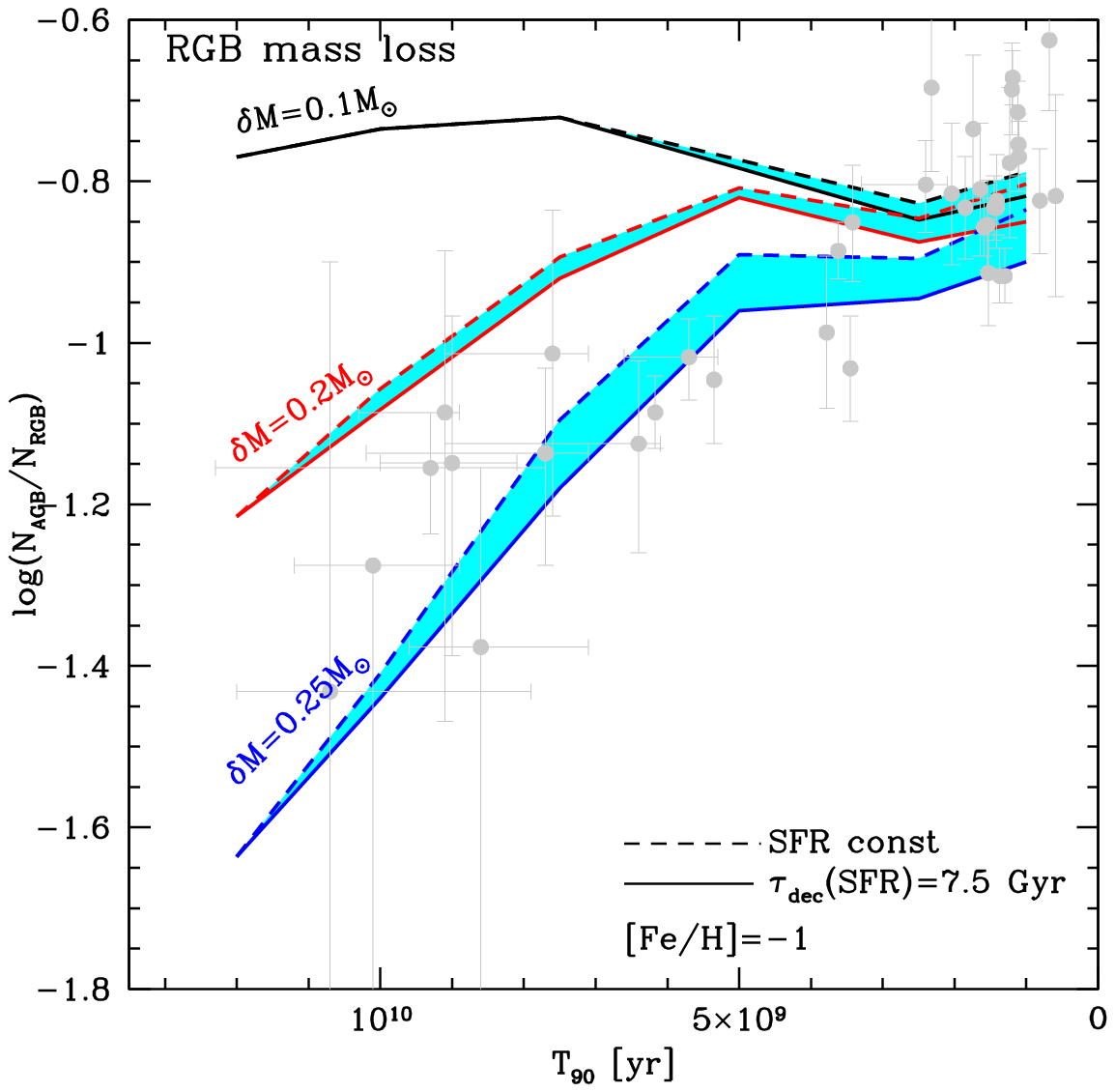}}
\end{minipage}
\begin{minipage}{0.48\textwidth}
\resizebox{1.\hsize}{!}{\includegraphics{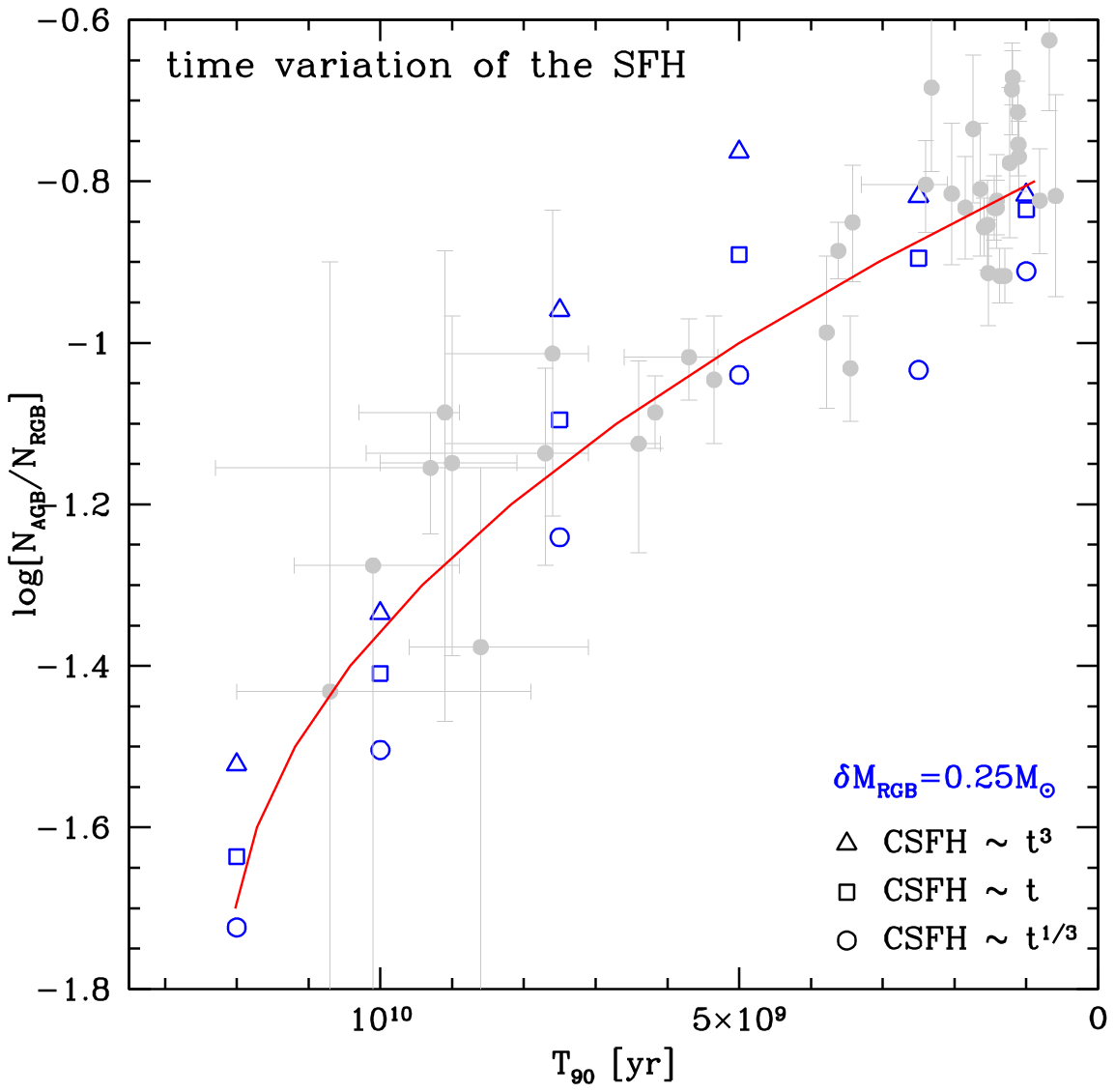}}
\end{minipage}
\vskip-90pt
\begin{minipage}{0.48\textwidth}
\resizebox{1.\hsize}{!}{\includegraphics{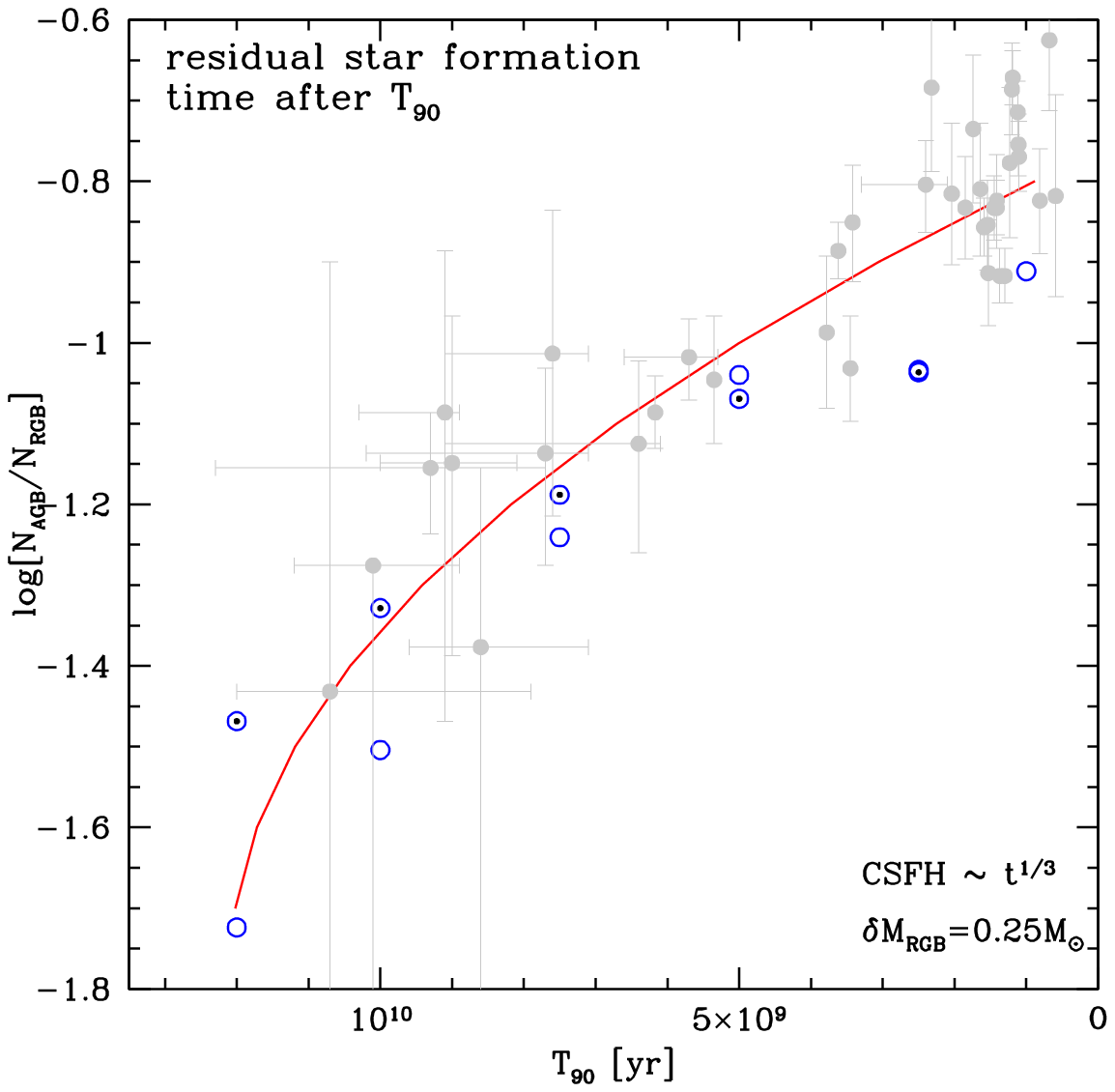}}
\end{minipage}
\begin{minipage}{0.48\textwidth}
\resizebox{1.\hsize}{!}{\includegraphics{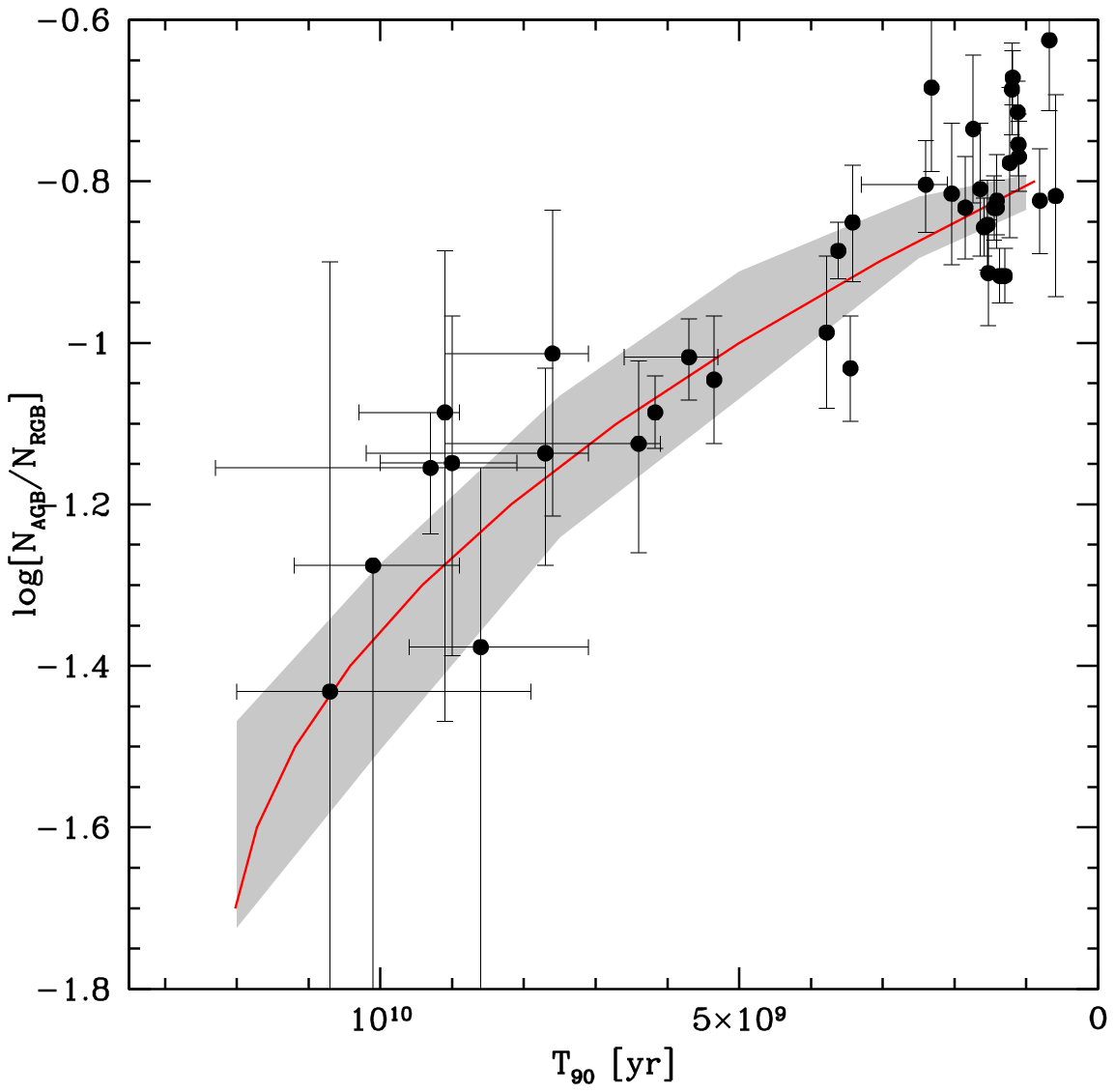}}
\end{minipage}
\vskip-60pt
\caption{$\rm N_{AGB}/N_{RGB}$ data of different galaxies are compared
with the predictions from population synthesis. The top, left panel
reports the comparison between the data and the results obtained 
by assuming different mass losses during the RGB evolution of low-mass stars, 
and considering the two cases of a SFH constant (dashed lines) and decaying with a
time scale of 7.5 Gyr (solid line). In the top, right panel the data 
points are compared with the results based on the $\rm \delta M_{RGB}=0.25~M_{\odot}$
assumption, for different SFH vs time relations (for which the
CSFH $\rm \propto t^{1/3}, t, t^3$ cases are considered). The bottom, left panel refers to 
$\rm \delta M_{RGB}=0.25~M_{\odot}$ and CSFH $\rm \propto t^{1/3}$, 
where the two cases that the residual $10\%$ of the stars are formed either within 
half a Gyr or over the time until 1 Gyr ago are indicated with open points filled 
with black dots and with open points, respectively. The bottom, right panel
shows the comparison between the data points and the most likely
T90 vs. $\rm N_{AGB}/N_{RGB}$ relation (see text for details); the
uncertainty associated with the latter trend is indicated with the grey-shaded area
}
\label{fsum}
\end{figure*}

To reach a more general understanding of how the $\rm N_{AGB}/N_{RGB}$
ratio depends on the SFH, we simulate the evolution of galaxies characterised 
by different $\rm T_{90}$'s, where we explore the role played by the following 
factors: a) the average mass loss experienced by stars of sub-solar mass during 
the RGB phase; b) the time variation of the star formation rate, since the 
formation of the galaxies, until the $\rm T_{90}$ epoch; c) the duration of
the time required to form the residual $10\%$ of the stellar mass of the galaxy, 
after the $\rm T_{90}$ epoch. For point a), we consider the cases explored in 
the previous section, namely $\rm \delta m_{RGB}=0.1, 0.2, 0.25~M_{\odot}$. 
For point b), we assume that the time variation
of the cumulative star formation was $\rm CSFR \propto t^3$,
$\rm \propto t$, $\rm \propto t^{1/3}$. This is to describe a wide
range of situations, ranging from a rapidly growing star formation,
strongly peaked towards the lowest masses ($\rm CSFR \propto t^{1/3}$),
to a slower process, favouring the formation of higher mass stars
($\rm CSFR \propto t^{3}$). For what concerns
point c), we compare the results obtained by changing
the duration of the time interval after the $\rm T_{90}$ epoch:
we considered a variety of cases, ranging from the situation that
the residual $10\%$ of stars formed within half Gyr after the
$\rm T_{90}$ epoch, to the case that star formation continued at
low rates, until now. The outcome of this exploration is
summarised in Fig.~\ref{fsum}, where the data of
Fig.~\ref{figdata} are also shown.

For $\rm T_{90} \geq 10$ Gyr the most relevant
quantity affecting $\rm N_{AGB}/N_{RGB}$
is the RGB mass loss, as the stellar population is dominated by 
low-mass stars, whose duration 
of the phase during which they evolve into the AGB box
is highly sensitive to $\rm \delta m_{RGB}$ (see top, left panel
of Fig.~\ref{f0820}). 
A non-negligible role in the determination 
of $\rm N_{AGB}/N_{RGB}$ for the galaxies with the oldest populations 
is also played by the time when the star formation process 
ends, after the $\rm T_{90}$ epoch, whose effects are of the order
of $\sim 0.2$ dex. The latter finding, at first surprising, 
is due to the fact that in galaxies where the majority of
the stars descend from old, low-mass progenitors, that barely
enter the AGB box, an extended period of star formation,
though limited to $10\%$ of the global stellar mass, is
sufficient to significantly increase the number of stars
in the AGB box, thus $\rm N_{AGB}/N_{RGB}$.

Inspection of Fig.~\ref{fsum} shows that the impact of the 
factors described in points a-c above becomes progressively
smaller towards the shorter $\rm T_{90}$'s, and that the
overall uncertainty on $\rm N_{AGB}/N_{RGB}$ found in the 
$\rm T_{90}=1$ Gyr case is within 0.15 dex. This is again
related to the less and less relevant role played 
by $\rm M<0.9~M_{\odot}$ stars in the galaxies with 
$\rm T_{90}<5$ Gyr, and the larger presence of
the higher mass counterparts: indeed for the latter stars 
the duration of the staying within the AGB box is not 
particularly sensitive to the input adopted.

Despite the results reported in Fig.~\ref{fsum} indicate a
significant spread in the $\rm T90$ vs $\rm N_{AGB}/N_{RGB}$
relation, particularly in the galaxies with the oldest
populations, we may still consider $\rm N_{AGB}/N_{RGB}$ as
a good indicator of $\rm T90$, based on the following arguments:
a) according to the results discussed in the previous section, 
supported by the recent analysis by \citet{tailo20}, we may safely 
consider that, on the average, low mass stars suffered
an RGB mass loss in excess of $\rm 0.2~M_{\odot}$; b) in the
galaxies where no recent star formation occurred, such as the cases 
of Sculptor and NGC 185 discussed in the previous section, the star 
formation rate decreases rapidly since the formation epoch, thus the 
time variation of CSFH is better reproduced by the 
$\rm CSFH \propto t^{1/3}$ relation; c) in the galaxies characterised 
by recent star formation, such as KK 77 and Fornax, star formation 
proceeds with more continuity, thus the $\rm CSFH \propto t$
relation is more suitable to describe the behaviour of the 
cumulative star formation history. 

Taking all these factors into account, we expect that most galaxies
follow the $\rm T90$ vs $\rm N_{AGB}/N_{RGB}$ trend indicated
with a red line in Fig.~\ref{fsum}, which can be approximated by 
the expression:

$$
\rm{T_{90}/Gyr}=-12\times \log^2(\rm N_{AGB}/N_{RGB})
-40\times \log(\rm N_{AGB}/N_{RGB})-25
$$

The grey-shaded area in the bottom, right panel of Fig.~\ref{fsum}
represents the uncertainty associated with the aforementioned trend,
which we see to be of the order of 1 Gyr for the galaxies with old
stellar populations, and progressively decreases to $\sim$ half a 
Gyr for the galaxies with recent star formation. These conclusions
hold for the $\rm [Fe/H]=-1$ chemical composition, and can be likely
safely extended to metallicities a factor $\sim 2$ different from 
$\rm [Fe/H]=-1$.

A linear $\rm T90$ vs $\rm N_{AGB}/N_{RGB}$ relation
could potentially reproduce the position of the majority of the
blue points in the middle and right panels of Fig.~\ref{fsum}:
however, such linear approximation would be consistent only
with the $\rm CSFH \propto t^{3}$ case, which, as discussed in
the point b) above, is definitively not the case for the galaxies 
with the oldest stellar populations.

\section{Discussion}
To understand how the present findings 
are sensitive to the details of stellar modelling and to the choice
of the filter considered for the number counts, we first compare the
ATON results with those based on different stellar evolution codes, then we
discuss the possibility of adopting near-IR filters instead of 
F814W.

\subsection{Comparison with other models}
\begin{figure}
\begin{minipage}{0.48\textwidth}
\resizebox{1.\hsize}{!}{\includegraphics{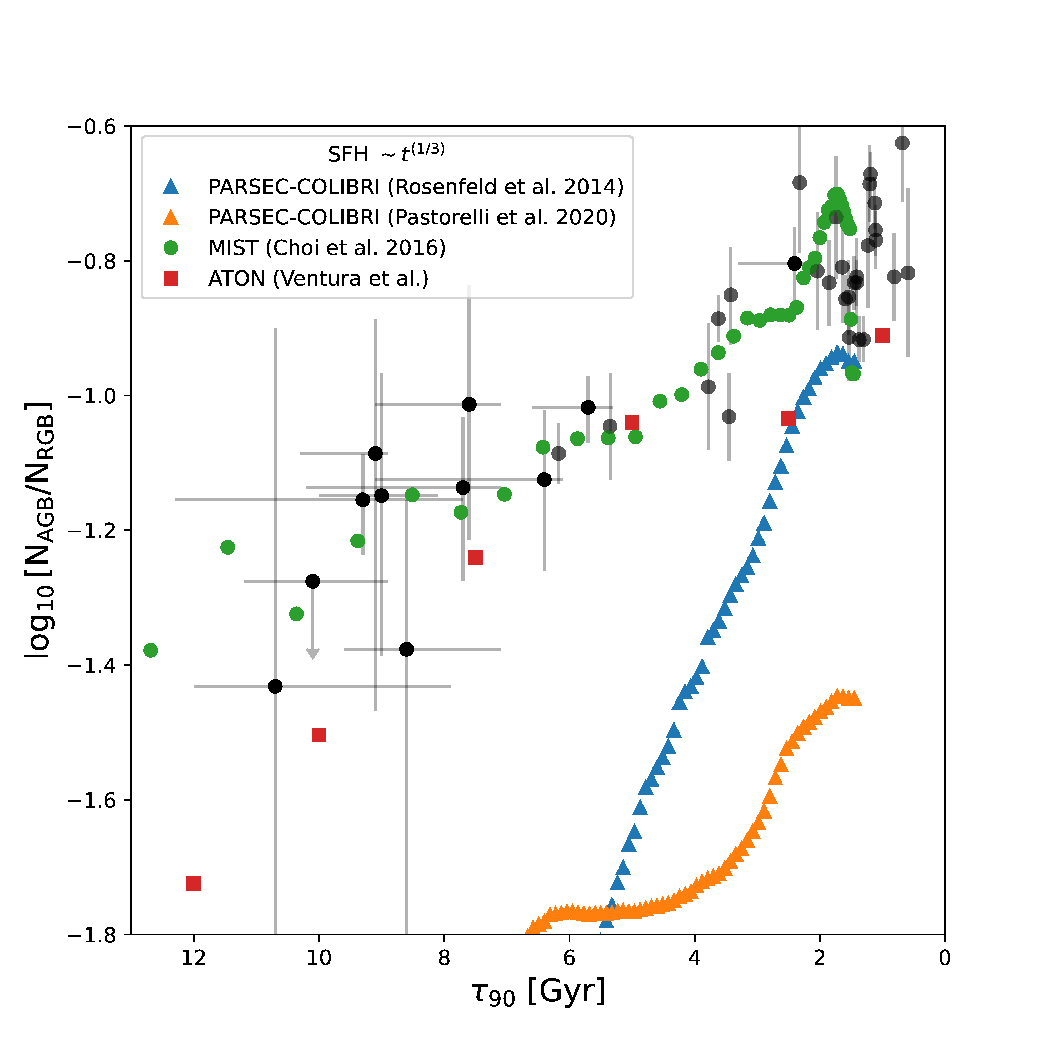}}
\end{minipage}
\caption{We compare our models with three others using a $\rm CSFH \propto t^{1/3}$ relation:  a) PARSEC-COLIBRI model of \cite{Rosenfeld2014}, b) PARSEC-COLIBRI model of \cite{Pastorelli2020}, c) MIST models of \citep{Choi2016}.
}
\label{fcomparison}
\end{figure}

In Fig. \ref{fcomparison} we 
compare the ATON $\rm T_{90}$ vs $\rm N_{AGB}/N_{RGB}$ relation 
with: a) the PARSEC-COLIBRI models by 
\citet{Rosenfeld2014}, b) the PARSEC-COLIBRI models by
\citet{Pastorelli2020}, and c) the MIST models by \citet{Choi2016}.
The PARSEC-COLIBRI models attain $\rm N_{AGB}/N_{RGB}$ values  
systematically lower than the ATON results and the data, especially 
in the most ancient epochs. The ATON and MIST results show more similarities, the differences 
being limited to the $\rm T_{90} > 8~Gyr$ epochs, when the 
$\rm N_{AGB}/N_{RGB}$ found when using the MIST results are $0.2-0.4$ dex larger than
those obtained in the ATON case; while the difference between the 
MIST and ATON models at younger ages $\sim1-2$ Gyr can be explained 
due to the difference in the treatment of dust in carbon-rich AGB stars.
At the oldest epochs the relevant factor distinguishing the various
computations is the expected number counts of stars in the 
AGB box: according to the computations presented in \citet{Pastorelli2020} 
the AGB box is not populated by stars older than $\sim 6$ Gyr, 
while the data suggests that this situation is found only for 
ages above 10 Gyr. On the other hand, use of the MIST models 
leads to a more numerous population of stars in the AGB box 
in comparison to the ATON ones.

\begin{figure}
\begin{minipage}{0.48\textwidth}
\resizebox{1.\hsize}{!}{\includegraphics{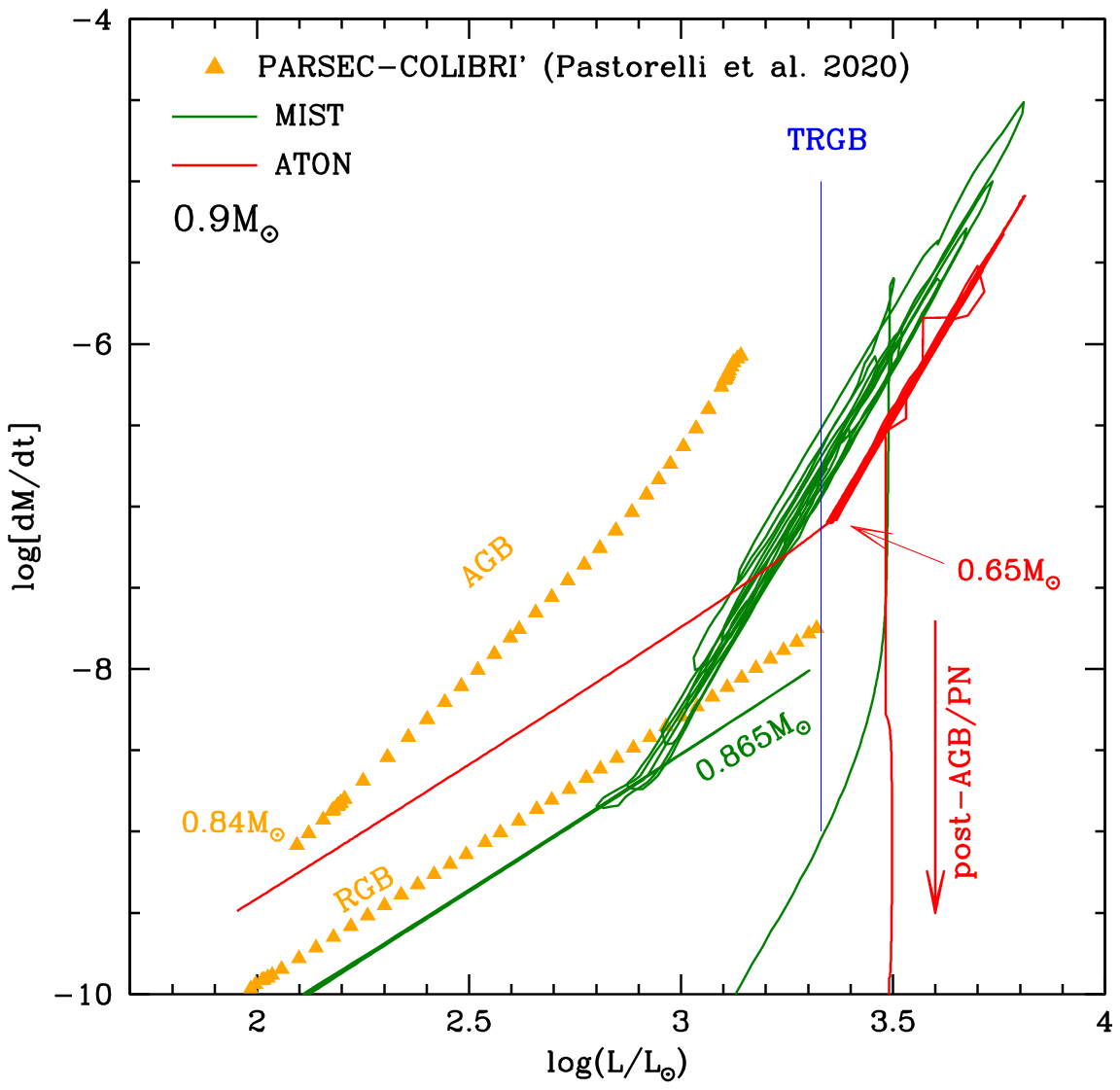}}
\end{minipage}
\vskip-60pt
\caption{Evolution of the mass-loss rate of a $\rm 0.9~M_{\odot}$
model star of metallicity $\rm [Fe/H]=-1$ as a function of the
stellar luminosity, across the RGB and the AGB phases. Orange
triangles indicate the evolution taken from \citet{Pastorelli2020},
while the green and red lines refer to the MIST and ATON
models, respectively. The masses reported along the evolutionary
tracks indicate 
the mass of the star at the ignition of the helium flash. The vertical, 
blue line indicates the location of the TRGB.
}
\label{fmloss}
\end{figure}

These differences are due to the treatment of mass loss
adopted in the various cases. We demonstrate this in 
Fig.~\ref{fmloss} where we show the evolution of the mass-loss 
rate during the RGB and AGB phases of a model star of 
initial mass $\rm 0.9~M_{\odot}$, which evolves on time 
scales of 7 Gyr.
We use the stellar luminosity as indicator of the evolutionary
phases, so the sequences must be followed rightwards until 
the TRGB, then must be resumed from 
the core helium burning.
The PARSEC AGB sequence (for clarity reasons we restrict to the
\citealt{Pastorelli2020} models, but similar results are found when
the \citealt{Rosenfeld2014} case is considered) is interrupted when the whole
envelope is lost, whereas for the MIST and ATON cases the
sequences are extended to the start of the post-AGB
evolution.

Inspection of Fig.~\ref{fmloss} shows that little mass loss occurs
during the RGB phase in the PARSEC case, with $\rm \dot{M} < 2\times 10^{-8}~M_{\odot}/$yr, 
so that only $\rm \sim 0.05~M_{\odot}$ are lost during the RGB. 
A similar result is found with the MIST models. In contrast, in the ATON 
case $\rm 0.25~M_{\odot}$ are lost during the RGB phase, so that the post-flash 
evolution starts with a lower mass than in MIST and PARSEC. At the ignition of 
the helium flash, the $\rm 0.9~M_{\odot}$
model star has a mass of $\rm 0.84~M_{\odot}$, $\rm 0.865~M_{\odot}$ and 
$\rm 0.65~M_{\odot}$  in the PARSEC, MIST and ATON models respectively.

During the post horizontal branch evolution, the mass-loss rates of the
ATON model preserve the consistency between the RGB and the E-AGB phases.
The net result is that the evolutionary tracks of the $\rm 0.9~M_{\odot}$
model star enter the AGB box, where it ultimately undergoes
TPs with mass-loss rates much higher than those in the RGB phase. 
However, the PARSEC mass-loss rates in the E-AGB phases, represented by the
upper series of triangles in Fig.~\ref{fmloss}, are much higher than those 
experienced during the RGB, attaining values close to $\rm 10^{-6}~M_{\odot}/$yr
at luminosities of the order of $\rm 1500~L_{\odot}$. Consequently, the 
evolutionary tracks of the stars do not enter the AGB box, because the 
whole envelope is lost before the luminosities reach the values attained 
at the TRGB. This is the reason for the extremely low 
$\rm N_{AGB}/N_{RGB}$'s found in the $\rm T_{90} > 4~Gyr$ epoch 
in the PARSEC cases.
If the same $\rm \dot M ~ vs ~ L$ trend  used in \citet{Pastorelli2020} 
to model the post-HB phases is applied to the RGB evolution, we find that 
low-mass stars never reach the tip of the RGB; thus even the RGB box 
would barely be populated. Since the surface gravities of RGB and post-HB 
stars of the same luminosities are similar, we believe that it is more
appropriate that the mass-loss rates adopted to model the two phases be similar,
rather than invoking that a sudden rise of $\rm \dot M$ takes place during 
the post-HB evolution.

The treatment of mass loss in the ATON and MIST cases show interesting similarities.
First, the mass loss rates experienced during the E-AGB phase are similar to 
those experienced during the RGB phase. Also the mass loss rates experienced
by the ATON and MIST models during the AGB phase 
are similar (see Fig.~\ref{fmloss}). As discussed previously, the main difference between the 
two models is the mass loss during the RGB phase, which reflects into the mass
with which the stars start the core helium burning phase, which is higher in the
MIST case: consequently the evolutionary tracks of the MIST models populate the AGB box 
for a longer time than in the ATON ones, thus favouring a slightly higher $\rm 
N_{AGB}/N_{RGB}$'s for ages $>7$ Gyr.

It is not possible to discriminate between the ATON and MIST descriptions
on the basis of the observational RGB and AGB number counts for galaxies 
of metallicity $\rm [Fe/H]=-1$: the differences are found only for those
systems in which star formation is limited to the ancient epochs, for which, 
unfortunately, the Poisson uncertainties in the observed $\rm N_{AGB}/N_{RGB}$ are 
so large that both models are consistent with the data (see Fig.~\ref{fcomparison}). 

To constrain mass loss, we need to turn to other phases of stellar evolution
where the differences in mass loss by low-mass stars have a more dramatic effect.
It has long been known that mass loss by low-mass stars during RGB evolution has
a strong impact on the morphology of the HB stars in globular clusters, which is
known to extend to the blue when a significant fraction of the
envelope is lost before the TRGB is reached, while it is clustered on the
red side of the colour-magnitude diagrams when the RGB mass 
loss is negligible. Several dedicated studies on the HB of globular clusters 
showed that low-mass, metal-poor stars, must undergo an RGB mass loss of 
at least $\rm 0.2~M_\odot$ in order to reproduce the observations of 
HB stars \citep{gratton10, pavel17, vand18, tailo20, tailo22, li25}: this is 
at odds with the MIST description, in which the RGB mass loss is below 
$\rm 0.05~M_{\odot}$, while being consistent with the ATON $\rm \delta m_{RGB}$ 
values derived in Section \ref{galaxies}. This suggests that it is important
that stellar models simultaneously model all these connected phases of stellar 
evolution, i.e., the RGB phase, the horizontal branch phase and the E-AGB/AGB phases,
in order to have a definitive handle on constraining mass loss in late-type stars.

\begin{figure*}
\begin{minipage}{0.33\textwidth}
\resizebox{1.\hsize}{!}{\includegraphics{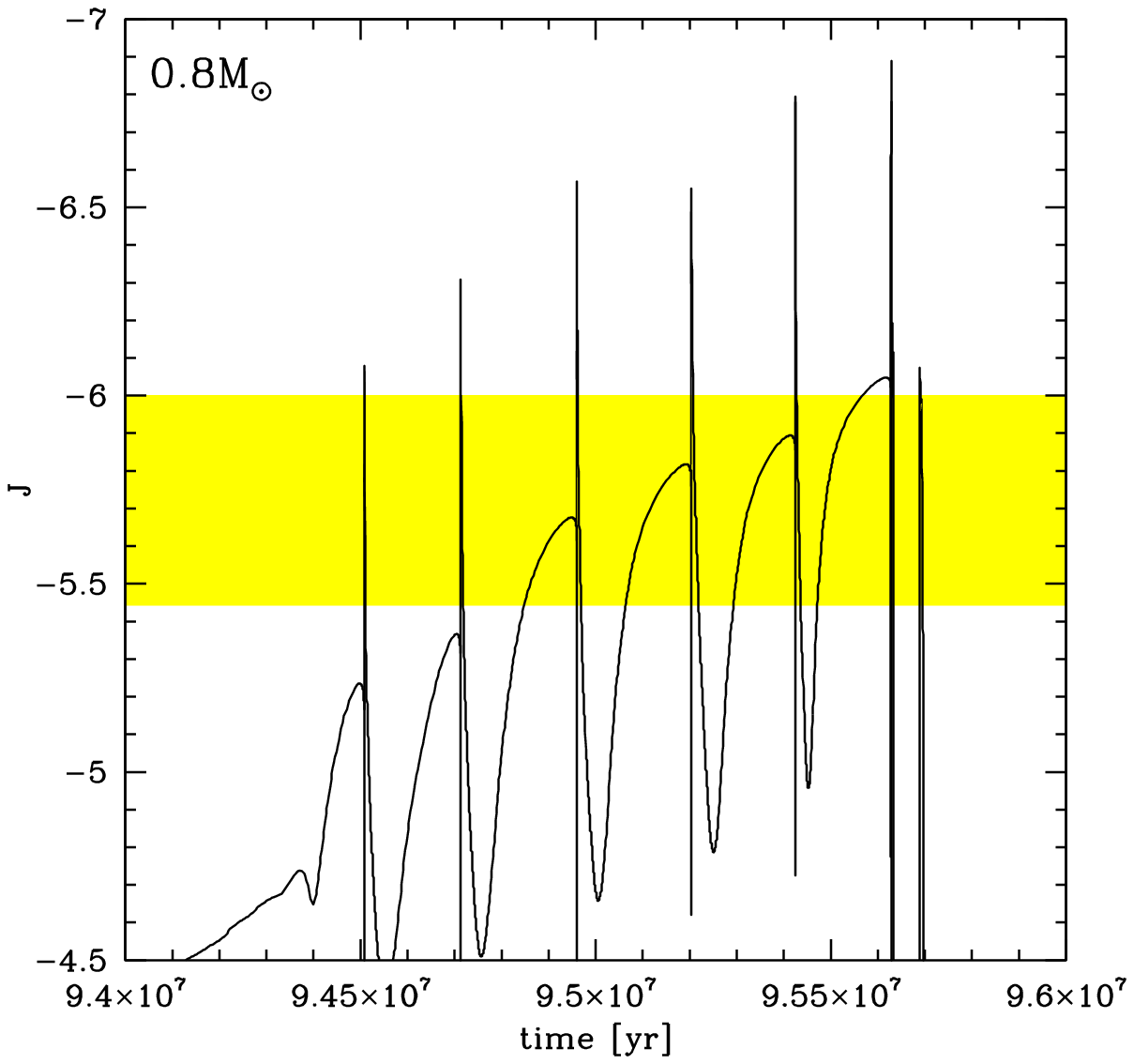}}
\end{minipage}
\begin{minipage}{0.33\textwidth}
\resizebox{1.\hsize}{!}{\includegraphics{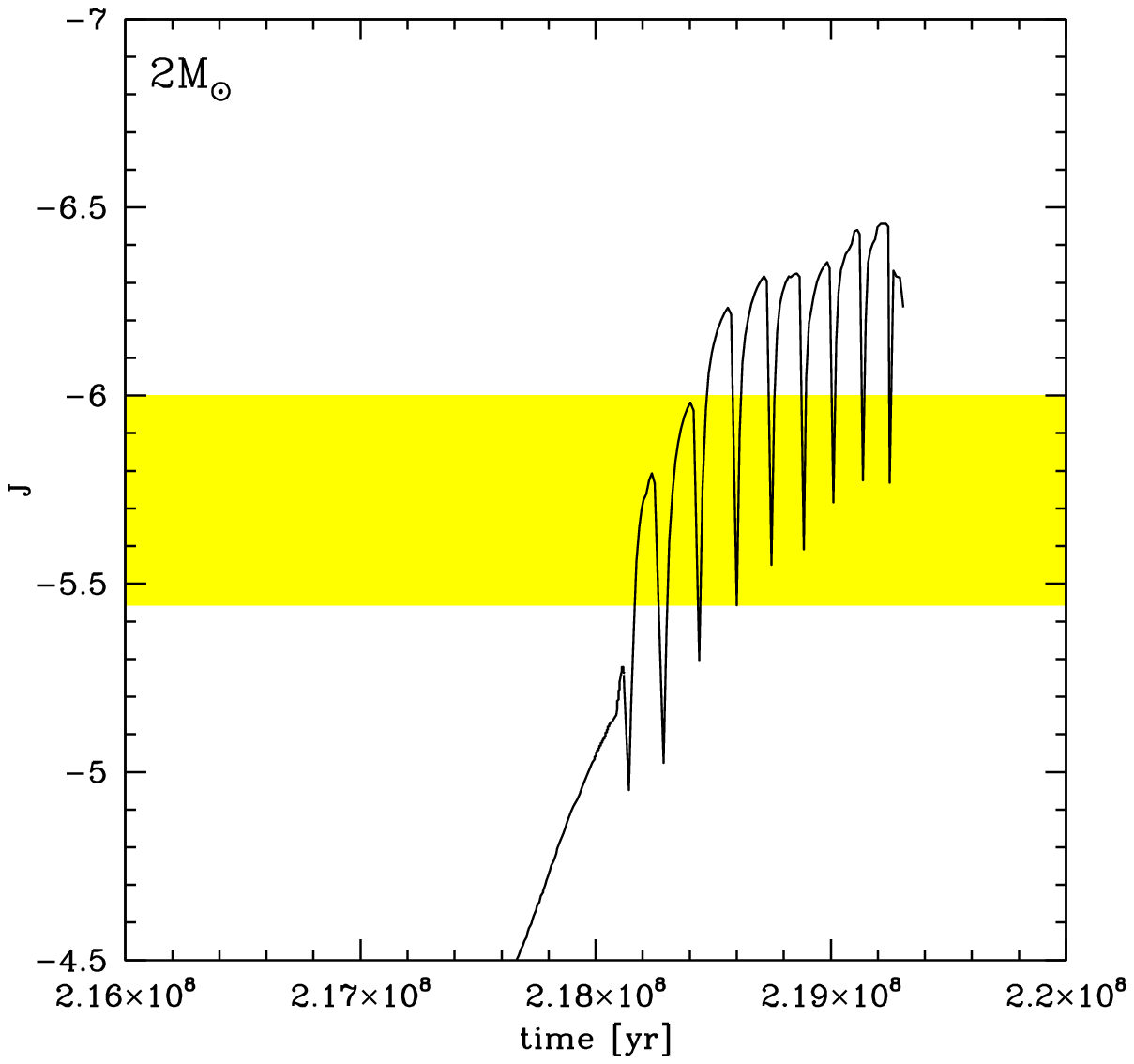}}
\end{minipage}
\begin{minipage}{0.33\textwidth}
\resizebox{1.\hsize}{!}{\includegraphics{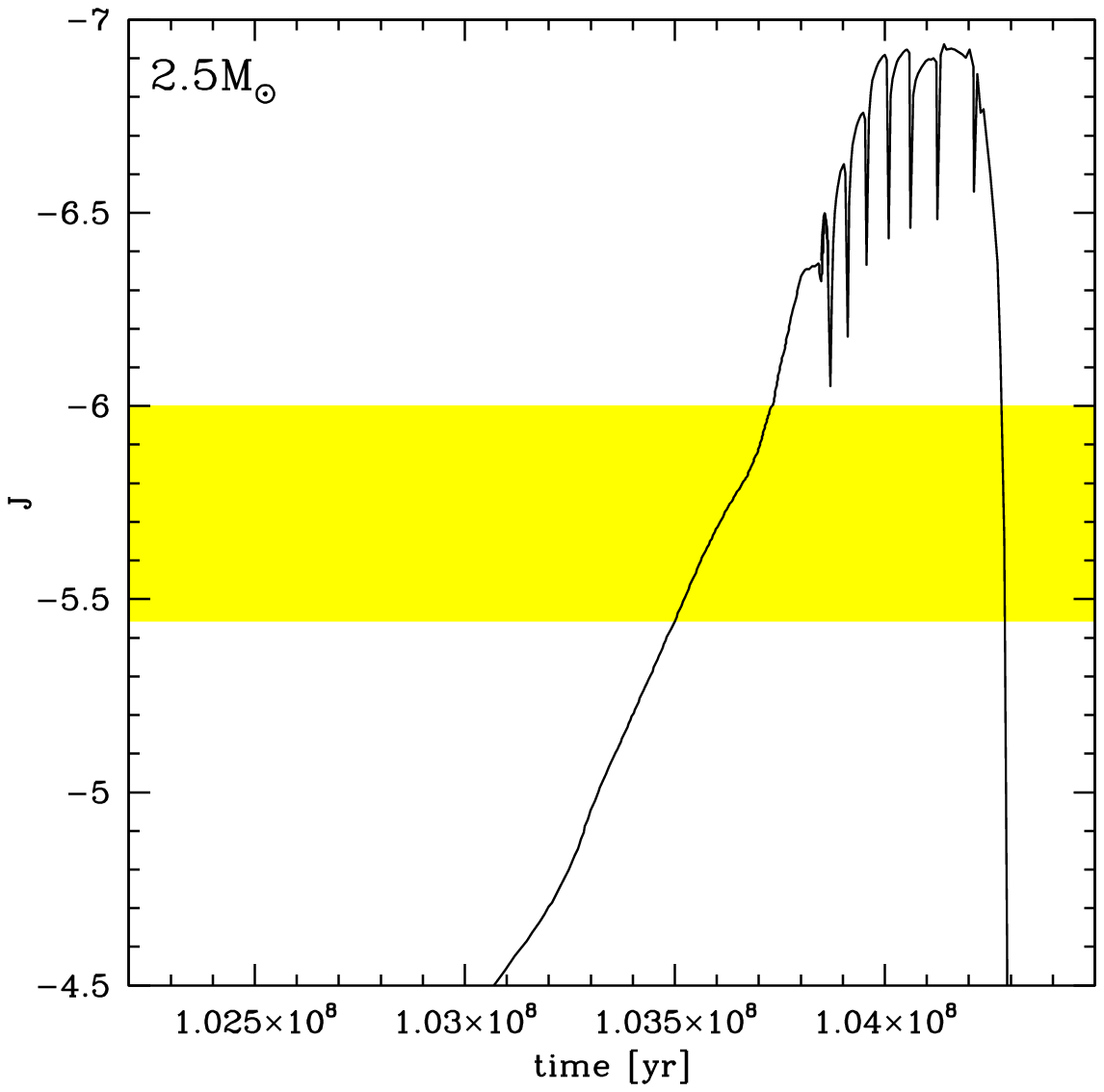}}
\end{minipage}
\vskip-40pt
\caption{Time variation of the $J$ magnitude of model 
stars of initial mass $\rm 0.8~M_{\odot}$ (left panel), $\rm 2~M_{\odot}$
(middle) and $\rm 2.5~M_{\odot}$ (right). For the $\rm 0.8~M_{\odot}$ model 
star only the case with no RGB mass loss is shown. The yellow-shaded
region brackets the $J$ magnitude range extended from 0.15 mag
to 0.75 mag above the TRGB.}
\label{figj}
\end{figure*}

\subsection{Extension to near-Infrared bands}
Since AGB and RGB stars are intrinsically cool and red, they can be clearly 
identified in the near-infrared bands. AGB stars are also less affected by
dust in the near-infrared bands. In light of ongoing and upcoming 
near-IR space missions such as JWST, Euclid, and the Roman Space Telescope, 
this becomes especially relevant. For this reason, we consider it worthwhile to 
investigate how the $\rm T90$ vs $\rm N_{AGB}/N_{RGB}$ relation behaves in the J filter 
and its analogues, where AGB and RGB observations will soon be available for hundreds 
of galaxies. Furthermore, infra-red colours involving the J filter can help separate M-giants 
from dwarfs \citep[e.g.][]{Majewski03}. In addition to the J filter (2MASS), in this 
subsection we also concentrate on the JWST F115W filter (which will be used to observe 
the faint stellar haloes of 28 galaxies out to 20 Mpc; Smercina et al. 2024; JWST 
proposal 8277), the Js filter of Euclid and the F129 filter of the Nancy Grace Roman 
Space telescope.

We begin by evaluating the possible choices of the RGB and AGB boxes in the
2MASS J filter. Fig.~\ref{figj} shows the time evolution of the J
magnitude of the same model stars reported in Fig.~\ref{f0820}.
The yellow-shaded region in the panels of Fig.~\ref{figj}
indicates a box extending over a J magnitude range 
equal to the one chosen by H23 to select the AGB box in the $\rm F814W$
band, i.e. from 0.15 mag to 0.75 mag above the TRGB.

The comparison between the results shown in Fig.~\ref{f0820} and those 
of Fig.~\ref{figj} outlines that while the range of $\rm F814W$
values reached during the AGB phase is roughly constant for the
stars of mass in the $\rm 0.8-2~M_{\odot}$ interval, the J fluxes
attained during this phase show up a higher sensitivity to the stellar mass: 
as an example, the $\rm 2~M_{\odot}$ model star, shown in the top, right
panel of Fig.~\ref{figj}, evolves to one J magnitude above the TRGB, 
whereas in the $\rm 0.8~M_{\odot}$ case, shown in the top left panel
of Fig.~\ref{figj}, this excursion is limited to 0.5 mag.
This behaviour is motivated by the change in the SED of the stars when
dust formation in the wind begins. The $\rm F814W$ is more
affected by the presence of dust in the circumstellar envelope:
despite the gradual increase in the luminosity, the progressive
shift of the SED of the stars to the IR spectral region
makes $\rm F814W$ decrease, after reaching a maximum during the
evolutionary phases preceding dust production (see Fig.~\ref{f0820}).
On the other hand, the J flux is less sensitive to the presence of
dust, particularly during the phases immediately following the start
of dust production, when the reprocessing of the radiation released by the
star is not very efficient: as shown in Fig.~\ref{figj}, in the stars
of mass $\rm M\leq 2~M_{\odot}$, the J flux continues to increase 
until the final AGB phases, thus reflecting more the 
variation of the luminosity of the star, in turn related to the stellar mass.

These arguments indicate the possibility of applying the same technique
proposed by H23 to the $\rm (J-K, J)$ diagram, as the latter, particularly
if an AGB box covering a mag interval higher than that proposed by H23 is used,
offers a better opportunity to trace the stars formation occurred during the last
$\sim 1.5$ Gyr, when $\rm M > 1.5~M_{\odot}$ stars evolve through the
AGB phase.


\begin{figure}
\begin{minipage}{0.48\textwidth}
\resizebox{1.\hsize}{!}{\includegraphics{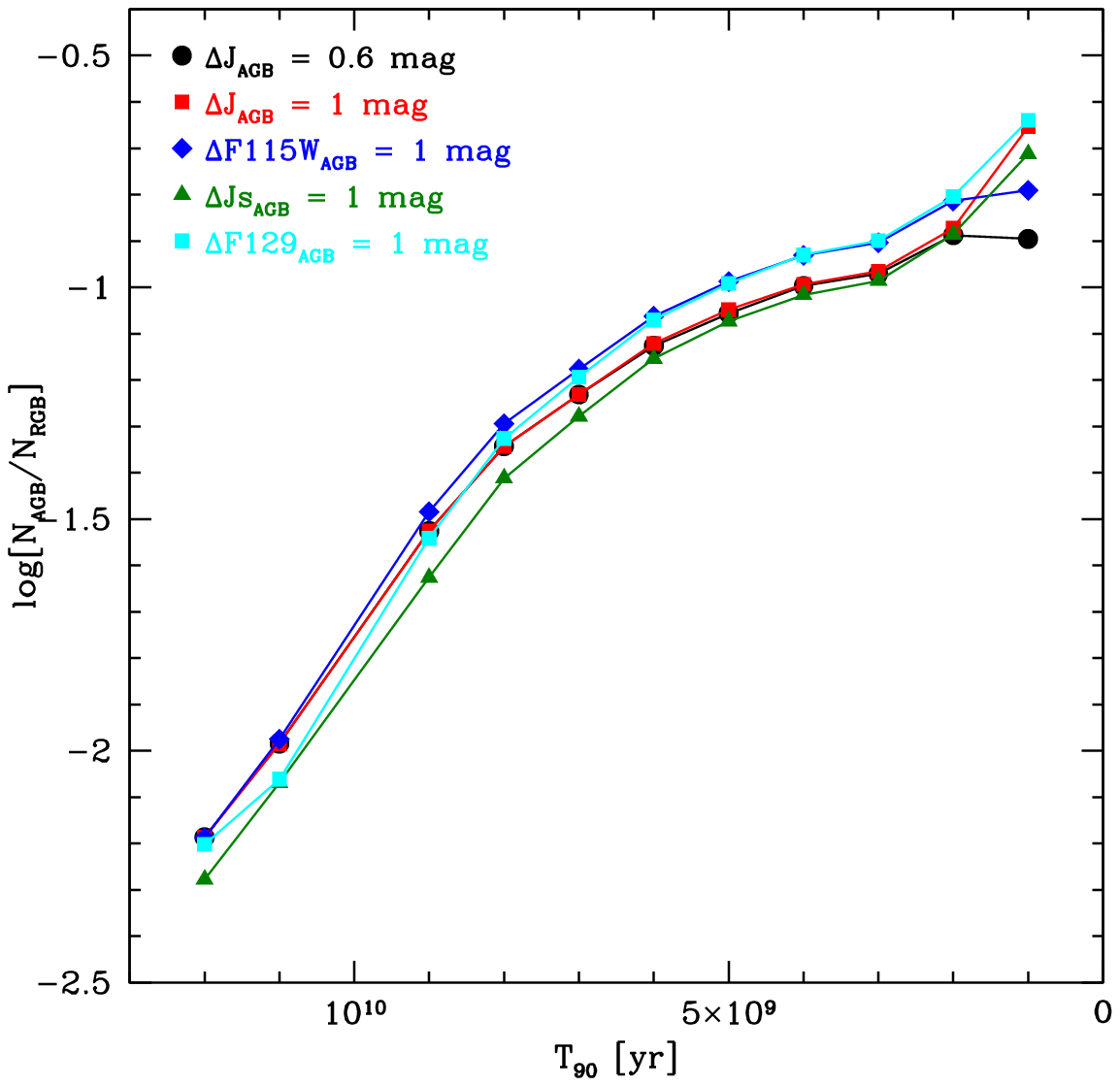}}
\end{minipage}
\vskip-60pt
\caption{The expected $\rm N_{AGB}/N_{RGB}$ as a function of T90,
where the RGB and AGB boxes are selected in the $J$, 
$\rm F115W$, $\rm J_s$, $\rm F129$ filters.
The results refer to the case of RGB mass loss 
$\rm \delta M_{RGB}=0.25~M_{\odot}$ and time variation of the
CSFR as $\rm t^{1/3}$. 
}
\label{figsumJ}
\end{figure}

Fig.~\ref{figsumJ} shows the trend between T90 vs $\rm N_{AGB}/N_{RGB}$ for 
various filters and AGB boxes. For this exercise, we assume that 
$\rm \delta M_{RGB}=0.25~M_{\odot}$ and a time variation of
the CSFH $\rm \sim t^{1/3}$ - consistent with the best agreement with
the data points shown in Fig.~\ref{figdata}. We first consider a standard 
AGB box with the same extension in the 2MASS J filter as that proposed 
by H23, namely 0.6 mag. We also consider a larger AGB box extending
over 1 mag. We find that the height of the AGB box has no effect 
on the estimated $\rm N_{AGB}/N_{RGB}$ for galaxies with T90 below 2 Gyr, 
whereas differences of the order of $\sim 0.2$ dex are found for galaxies 
with more recent star formation.

The same arguments hold when the F115W filter of JWST, the Js filter of
Euclid and the F129 filter of the Roman telescope are cosnidered. Indeed,
we see in Fig.~\ref{figsumJ} that the $\rm N_{AGB}/N_{RGB}$ vs. T90 trends derived 
when AGB boxes extending over 1 mag in J, F115W, Js and F129 are similar.

\subsection{Towards a full star formation history of a stellar population}
For galactic systems of metallicity $\rm [Fe/H]=-1$, the limited number of 
AGB and RGB stars allows us to constrain only a single parameter ($\mathrm{T}_{90}$) 
of the SFH instead of its full parametric form. On the other hand, galactic systems
at higher metallicities (e.g., $\rm [Fe/H]]= 0.0$) have much larger stellar masses. 
With an abundance of AGB and RGB stars, it is possible to attempt to constrain the 
parametric form of the SFH \citep[e.g.,][]{Rejkuba2022,Lee2025} of a stellar population. 
However, the principal uncertainty of the AGB models used to constrain the SFH is 
the mass loss at these higher metallicities. Although RGB mass loss has long been thought to 
be an increasing function of metallicity \citep{gratton10, pavel17, vand18, tailo20, tailo22}, 
there is recent evidence that at solar metallicity the trend may reverse 
\citep{Brogaard2024, Lee2025}. Hence, instead of blindly extrapolating mass loss to 
higher metallicities, it is essential to empirically verify the mass loss before moving 
on to using AGB stars to constrain the full SFH.


\section{Conclusions}
\label{concl}
Results from stellar evolution modelling combined with the description of
the dust formation process in the wind of evolved stars were used to
investigate the relationship between the duration of the star formation
process in galaxies, specifically T90, the time within which $90\%$ of the
stars formed, and the relative fractions of stars populating two selected
regions of the $\rm (F606W-F814W, F814W)$ diagram, suggested by H23: an RGB 
box, in the region just below the TRGB, and an AGB box, located in the brighter
part of the diagram.

The estimated $\rm N_{AGB}/N_{RGB}$ is critically sensitive to  
the mass lost by low-mass stars
during the ascent of the RGB, which affects
the duration of the AGB phase, and the possibility that the
stars evolve into the AGB box. The time variation of the SFH is also
relevant for the determination of $\rm N_{AGB}/N_{RGB}$. The
comparison with the data of the galaxies in the Local Group
with known T90 and $\rm N_{AGB}/N_{RGB}$ indicates
that an average RGB mass loss of $\rm 0.2-0.25~M_{\odot}$ 
occurred, consistent with independent measures of mass loss along 
the RGB from the colors of HB stars. Regarding the star formation 
process, the best agreement with the data points is obtained with the assumption that
the SFH either was strongly peaked towards the ancient epochs, 
which is the case for the galaxies in which star formation 
was halted within a few Gyr, or it kept approximately constant,
in the galaxies with more recent star formation. Overall, from 
this analysis it is possible to deduce a T90 vs 
$\rm N_{AGB}/N_{RGB}$ trend, which can be used to infer the
epoch when star formation stopped.

The present methodology was shown to work even better in the 
$\rm (J-K, J)$ and $\rm (F115W-F277W, F115W)$ diagrams, and also
in the diagrams constructed with the near-IR filters mounted
onboard of Euclid and the Roman Space Telescopes. Indeed 
the AGB variation of the J, F115W, Js and F129 magnitudes are more
sensitive to the initial mass of the star than F814W, which
offers the possibility of gaining a deeper understanding of the
star formation process in the recent epochs, provided that
a taller AGB box than that proposed by H23 for F814W is adopted.

\begin{acknowledgements}
      PV acknowledges support by the INAF-Theory-GRANT 2022 "Understanding mass loss and dust production from evolved stars"
\end{acknowledgements}

%
%

\begin{appendix}
\section{Observational Data}
\label{appendix:a1} 

{Here, we updated the list of galaxies that were used to calibrate the relationship between AGB/RGB ratio and $t_{90}$ in Table \ref{Obs_calibration} which was initially presented in H23. In particular, we recalculated the AGB/RGB ratio of NGC 147 and NGC 185 from the data of \cite{geha15}, while we updated and expanded the list of dwarf galaxies of Andromeda from the sample of \cite{savino25}. Furthermore, from the galaxies from the ANGST survey \citep{Dalcanton2009} presented in H23, we removed the galaxy NGC 4163 due to large uncertainties in its SFH.}

\begingroup
\renewcommand{\arraystretch}{1.4}
\begin{table}\centering
    \begin{tabular}{lccl}
    
    Galaxy      &AGB/RGB       &$t_{90}$/Gyr & References \\
    \hline 
    Fornax & $0.157^{+0.021}_{-0.020}$ & $2.4^{+0.9}_{-0.3}$ & 1, 2 \\
    NGC 147 & $0.096^{+0.011}_{-0.011}$ & $5.7^{+0.9}_{-0.4}$ & 8 \\
    NGC 185 & $0.070^{+0.012}_{-0.012}$ & $9.3^{+3.-}_{-1.6}$ & 8 \\
    Cassiopeia III & $0.082^{+0.048}_{-0.048}$ & $4.1^{+2.5}_{-1.5}$ & 9 \\
    Andromeda II & $0.097^{+0.049}_{-0.036}$ & $7.6^{+1.5}_{-0.5}$ & 4, 9 \\
    Andromeda I & $0.071^{+0.037}_{-0.030}$ & $9.0^{+1.0}_{-0.9}$ & 4, 9 \\
    Andromeda VI & $0.075^{+0.020}_{-0.020}$ & $6.4^{+2.7}_{-0.3}$ & 9 \\
    Andromeda VII & $0.073^{+0.020}_{-0.020}$ & $7.7^{+2.5}_{-0.6}$ & 9 \\
    Lacerta I & $0.042^{+0.028}_{-0.028}$ & $8.6^{+1.0}_{-1.5}$ & 9 \\
    Andromeda III & $<0.053$ & $10.1^{+1.1}_{-1.2}$ & 4, 9 \\
    Sculptor & $0.037^{+0.089}_{-0.037}$ & $10.7^{+1.3}_{-2.8}$ & 5, 2 \\
    Antlia &  $0.207^{+0.056}_{-0.044}$ & 2.32 & 6, 7 \\
    FM1/F6D1 & $ 0.090^{+ 0.018}_{-0.015}$ & 5.35 & 6, 7 \\
    Sc22/Sc-dE1 & $0.141^{+0.025}_{-0.022}$ & 3.42 & 6, 7 \\
    IKN & $ 0.082 ^{+0.009}_{-0.008}$ & 6.17 & 6, 7 \\ 
    ESO 294-010 & $0.103^{+0.025}_{-0.020}$ & 3.78 & 6, 7 \\ 
    ESO 540-032 & $0.093^{+0.015}_{-0.013}$ & 3.45 & 6, 7 \\
    KDG 2 & $0.237^{+0.052}_{-0.043}$ & 0.68 & 6, 7 \\ 
    KK 77 & $0.130^{+0.011}_{-0.010}$ & 3.62 & 6, 7 \\
    ESO 410-005 & $0.147^{+0.023}_{-0.020}$ & 1.85 & 6, 7 \\
    HS 117 & $0.150^{ +0.024}_{-0.021}$ & 0.811 & 6, 7 \\
    KDG 63 & $0.139^{+0.012}_{-0.011}$ & 1.59 & 6, 7 \\ 
    UGC 8833 & $0.206^{+0.029}_{-0.025}$ & 1.20 & 6, 7 \\
    KDG 64 & $0.147^{+0.012}_{-0.011}$ & 1.41 & 6, 7 \\
    KDG 61 & $0.121^{+0.010}_{-0.009}$ & 1.30 & 6, 7 \\
    DDO 181 & $0.193^{+0.018}_{-0.016}$ & 1.12 & 6, 7 \\
    KKH 98 & $0.167^{+0.040}_{-0.032}$ & 1.23 & 6, 7 \\
    KDG 73 & $0.184^{+0.043}_{-0.035}$ & 1.74 & 6, 7 \\
    KKH 37 & $0.140^{+0.019}_{-0.017}$ & 1.54 & 6, 7 \\
    UGCA 292 & $0.152^{+0.051}_{-0.038}$ & 0.59 & 6, 7 \\
    DDO 113 & $0.122^{+0.020}_{-0.017}$ & 1.53 & 6, 7 \\
    DDO 44 & $0.147^{+0.014}_{-0.013}$ & 1.45 & 6, 7 \\
    GR8 & $0.155^{+0.032}_{-0.027}$ & 1.64 & 6, 7 \\
    DDO 78 & $0.121^{+0.010}_{-0.009}$ & 1.37 & 6, 7 \\
    DDO 6 & $0.150^{+0.021}_{-0.019}$ & 1.41 & 6, 7 \\
    UGC 8508 & $0.176^{+0.016}_{-0.015}$ & 1.11 & 6, 7 \\
    NGC 3741 & $0.170^{+0.018}_{-0.016}$ & 1.104 & 6, 7 \\ 
    KDG 52 & $0.153^{+0.034}_{-0.028}$ & 2.04 & 8, 10 \\
    DDO 53 & $0.213^{+0.017}_{-0.016}$ & 1.19 & 6, 7 \\
    \hline 
    \end{tabular}
    \caption{AGB/RGB ratios and $t_{90}$ values for the observational sample of calibrating galaxies. References: 1) \citet{deBoer2012}, 2) \citet{Weisz2015}, 3) Hubble Source Catalog photometry WFC3 F606W/F814W (GO-15336; PI: A. Ferguson), 4) \citet{Skillman2017}, 5) \citet{deBoer2011}, 6) \citet{Dalcanton2009}, 7) \citet{Weisz2011}, 
    8) \citet{geha15}, 9) \citet{savino25}}
    \label{Obs_calibration}
\end{table}
\endgroup
\end{appendix}

\end{document}